\documentclass[apj]{emulateapj}
\usepackage{graphicx}
\usepackage{times}
\usepackage{amsmath}
\usepackage{color}
\usepackage{graphics}
\usepackage{subfigure}
\usepackage{txfonts}
\usepackage{natbib}
\usepackage{wasysym}
\usepackage{afterpage}
\usepackage[flushleft]{threeparttable}

\bibpunct{(}{)}{;}{a}{}{,}

\newcommand{\BE}{\begin{equation}}
\newcommand{\EE}{\end{equation}}
\newcommand{\BA}{\begin{eqnarray}}
\newcommand{\EA}{\end{eqnarray}}

\newcommand{\rmd}{{\rm d}}

\newcommand{\Nabla}{\vec{\nabla}}

\newcommand{\dV}{\, \rmd \mathcal{V}}
\newcommand{\surf}{{\partial \mathcal{V}}}
\renewcommand{\vec}[1]{ {\mathbf #1} }
\newcommand{\vol}{\mathcal{V}}

\newcommand{\vA}{\vec{A}}

\newcommand{\vAp}{\vA_{\rm p}}

\newcommand{\vB}{\vec{B}}

\newcommand{\vBp}{\vB_{\rm p}}

\newcommand{\vBj}{\vec{B}_{\rm j}}

\newcommand{\curlAp}{\Nabla \times \vAp}  
\newcommand{\E}{E_{\rm V}}             %total energy 
\newcommand{\Ep}{E_{\rm p}}      %energy of the potential field
\newcommand{\Ej}{E_{\rm j}}      %energy of the current-carrying field
\newcommand{\Ens}{E_{\rm ns}}

\newcommand{\Hj}{H_{\rm j}}
\newcommand{\Hpj}{H_{\rm pj}}
\newcommand{\Hv}{H_{\rm V}}

\shorttitle{Non-potential magnetic helicity ratios at the onset of eruptions}
\shortauthors{F.~P.~Zuccarello et al.}

\begin{document}

\title{Threshold of non-potential magnetic helicity ratios at the onset of solar eruptions}

\author{Francesco P. Zuccarello\altaffilmark{1,2}}\email{Francesco.Zuccarello@kuleuven.be}
\author{E.~Pariat\altaffilmark{2}}\email{Etienne.Pariat@obspm.fr}
\author{G.~Valori\altaffilmark{3}}\email{g.valori@ucl.ac.uk}
\author{L.~Linan\altaffilmark{2}}\email{Luis.Linan@obspm.fr}

\altaffiltext{1} {Centre for mathematical Plasma Astrophysics, Department of Mathematics, KU Leuven, Celestijnenlaan 200B, B-3001 Leuven, Belgium}
\altaffiltext{2} {LESIA, Observatoire de Paris, Universit\'{e} PSL, CNRS, Sorbonne Universit\'{e}, Univ. Paris Diderot, Sorbonne Paris Cit\'{e}, 5 place Jules Janssen, 92195 Meudon, France}
\altaffiltext{3} {UCL-Mullard Space Science Laboratory, Holmbury St. Mary, Dorking, Surrey, RH5 6NT, UK }

\begin{abstract}
 
The relative magnetic helicity is a quantity that is often used to describe the level of entanglement of non-isolated  magnetic fields, such as the magnetic field of solar active regions.
The aim of this paper is to investigate how different kinds of photospheric boundary flows accumulate relative magnetic helicity in the corona and if and how-well magnetic helicity related quantities identify the onset of an eruption.
We use a series of three-dimensional, parametric magnetohydrodynamic  simulations of the formation and eruption of magnetic flux ropes. All the simulations are performed on the same grid, using the same parameters, but they are characterized by different driving photospheric flows, i.e., shearing, convergence, stretching, peripheral- and central- dispersion flows. For each of the simulations, the instant of the onset of the eruption is carefully identified by using a series of relaxation runs. 
We find that  magnetic energy and total relative helicity  are mostly injected when shearing flows are applied at the boundary, while the magnetic energy and helicity associated with the coronal electric currents increase regardless of the kind of photospheric flows. We also find that, at the onset of the eruptions, the ratio between the non-potential magnetic helicity and the total relative magnetic helicity has the same value for all the simulations, suggesting the existence of a threshold in this quantity. Such threshold is not observed for other  quantities as, for example, those related to the magnetic energy.

\end{abstract}

\section{Introduction}

Over the last few years, the study of magnetic helicity, a quantity estimating the level of twist and entanglement of the magnetic field lines in a magnetised plasma, has received  renewed attention in solar physics. This evolution is enabled thanks to the development of several new methods to compute and represent magnetic helicity \citep{Rudenko11,Thalmann11,Valori12,YangS13, Dalmasse14,Yeates13,Yeates14,Prior14}. Among these new approaches, some allows to properly compute magnetic helicity in non-magnetically-isolated domains, i.e., in typical condition for natural plasma where the magnetic field is threading the boundaries of the studied domain \citep[see][ for a complete review and a benchmark of these methods]{Valori16}.  These new techniques now permit an exact and controlled estimation of magnetic helicity in three-dimensional datasets, and have in particular been applied to the study of the evolution of helicity in several numerical simulations of solar active events \citep{Moraitis14,Pariat15b,Pariat17,Sturrock15,Sturrock16} as well as in coronal magnetic field extrapolations of observed active regions \citep{Valori13,Moraitis14,GuoY17,James18}.

Magnetic helicity has recently been used as an innovative tool to study and better understand typical problems in solar physics such as the magnetic reconnection mechanism \citep{Russell15}, the formation of filament channels \citep{ZhaoL15,Knizhnik15,Knizhnik17} and their large scale distribution over the solar cycle \citep{Yeates16}, the solar dynamo \citep{Miesch16,Brandenburg17}, the formation of active regions \citep{LiuY12,LiuY14a,LiuY14b,Moraitis14,Pariat17}, the rotation of sunspots \citep{Sturrock15,Sturrock16}, and the generation of solar jets \citep{Karpen17}. 

A field of research in which magnetic helicity is expected to bring key results is the study of solar flare and eruptions. Even though magnetic helicity is only a strict invariant in ideal magnetohydrodynamic (MHD), \citet{Pariat15b} have confirmed \cite{Berger84b} scaling argument that helicity is quasi-conserved in active events even when intense non-ideal processes such as magnetic reconnection is acting to transform most of the magnetic energy. This now-demonstrated conservation of magnetic helicity is a key concept which is believed to be a ruling principle beyond the existence and the formation of coronal mass ejections \citep{Rust94,Low96,Green02,Mandrini05,Priest16}.

The study of the relationship between flare/eruptions and the magnetic evolution of active regions has been particularly prolific \citep[e.g.][]{Nindos02,Nindos04,ParkSH08,ParkSH10,ParkSH12,Tziotziou12,Tziotziou13,Tziotziou14, Zuccarello11,Zuccarello14, Zuccarello17}. So far most of the work has relied on computing magnetic helicity from observed series of magnetograms and estimating the helicity flux following the ground breaking method of \citet{Chae01}, which has, however, some  inherent limitations  \citep{Demoulin09}. It is nonetheless worth mentioning that several such observational studies have concluded on a close relation between high helicity content and enhance eruptivity \citep[e.g.][]{Nindos04,Labonte07,Smyrli10,Tziotziou12}.

The new and exact methods to compute helicity in a 3D domain are however now enabling the comprehensive study of magnetic helicity in numerical datasets. \citet{Pariat17} have recently studied parametric simulations of the formation of solar active regions leading either to stable configurations or to eruptions \citep[presented in][]{Leake13b,Leake14a}. They found that magnetic helicity was strongly discriminating between the different simulations. Furthermore, they showed that by using the helicity decomposition introduced by \citet{Berger03}, the ratio of the magnetic helicity of the current carrying part of the field to the total helicity could be used as a clear predictor of the eruptivity in the simulations. This quantity indeed presented high values only for the eruptive simulations  and only before the eruption. Additionally, this helicity ratio was no longer differentiating the eruptive simulation from the non-eruptive one after the eruption, when the system was stable in all the different runs.

The experimental set-up of \citet{Leake13b,Leake14a} does not permit to determine the existence of an eruptivity threshold related to the helicity ratio. The stability of the magnetic system was indeed likely deterministically given by the initial condition, i.e., for the eruptive simulations, the system was not brought from a equilibrium stage towards instability by controlled imposed quasi-steady forcing. Therefore, while remarkable, the results of \citet{Pariat17} was not conclusive on the reason why the helicity of the current carrying part of the field could be related to  an enhanced eruptivity. In order to determine whether their results was due to pure hazard or is symptomatic of a deeper physical meaning, the present study investigates the energy and helicity content of a radically different set of parametric simulations of eruptive events. This manuscript focuses on the analysis of the line-tied 3D MHD simulations of \citet{Zuccarello15}. In these simulations, eruptions are triggered by boundary-driven motions that mimics the long-term evolution of solar active regions, with the presence of shearing motions and large scale diffusion of the magnetic polarities. Unlike with the flux-emergence simulations of \cite{Leake13b,Leake14a}, the trigger time and mechanism have been carefully determined thanks to numerous relaxation runs. \citet{Zuccarello15} have shown that the eruptions where tightly related with the torus instability mechanism \citep[][]{Kliem06,Torok2007,Aulanier10,Demoulin10,Olmedo10,Kliem14}. 
A goal of the present study is to determine whether an helicity based eruptivity predictor is also able to describe the eruptivity stage of the simulations of \citet{Zuccarello15} and how it relates with the torus instability.

Additionally the parametric simulations of \citet{Zuccarello15} present different type of boundary driving motions. Thanks to the comparison of these different simulations it is possible to compare and determine which boundary motions are the most efficient at injecting total helicity in the coronal domain, as well as in the different terms  of the helicity decomposition. Helicity accumulation is indeed a fundamental process of the formation and evolution of active regions \citep{Green02,Green03,Mandrini05,LiuY12,LiuY14a,LiuY14b,Romano14,vanDriel15,Sturrock15,Sturrock16}. Studying the most efficient way by which  helicity is injected in active regions can reveal to be particularly important for determining their eruptivity potential.

The manuscript is organized as follows. The simulation setups and evolution of the system are discussed in Section~\ref{Sec:MHDsim}. The different magnetic energy and helicity decompositions  are presented in Section~\ref{Sec:Helicity}. Section~\ref{Sec:MagFlux} describes the evolution of the magnetic flux as a result of the applied boundary motions. Sections~\ref{Sec:Trends} and \ref{Sec:Thresholds} present the results of our analysis, i.e., the time evolution of the different magnetic energy and helicity decompositions and their values at the onset of the eruptions. Finally, in Section~\ref{Sec:Conclusion} we discuss our results and conclude.

\section{The MHD simulations}
\label{Sec:MHDsim}

To study the evolution of magnetic energy and helicity during the formation and eruption of magnetic flux ropes we solve the full three-dimensional MHD equations using the OHM-MPI code \citep{Aulanier05, Zuccarello15}. In this paper we analyze the same runs presented in \cite{Zuccarello15} where the MHD equations are solved in a non-uniform Cartesian grid that expands from the location $x=y=z=0$ and covers the  domain $x \in [-10, 10], ~y \in [-10, 10],~z \in [0, 30]$ where $x$ and $y$ are the horizontal directions and $z$ is the vertical one. The goal of that study was to carefully determine and investigate the onset of the eruptions in the framework of the torus instability. To achieve this goal, a parametric study consisting of four different simulations was performed.  For each of the four different simulations the time of the onset of the eruption was carefully determined using a series of relaxation runs.

The four simulations  share the same initial phase where the magnetic field is modified from an initial, potential configuration into a sheared one  (Sections~
\ref{Sec:InCon} and \ref{Sec:Shearing}). From this point onward,  four different boundary motions that result in four different ways to build a flux rope and bring it to the eruption point are applied (Section~\ref{Sec:Driving}). Finally, a proper eruption phase follows in each of the four runs (Section~\ref{Sec:Erupt}). The first two phases, shearing and flux rope formation, are the most relevant ones for the study of helicity evolution discussed in this article.

%-----------------------------------------------------------------------------------
% Shearing Figure
%-----------------------------------------------------------------------------------

\begin{figure*}
\begin{center}
\subfigure{
\includegraphics[width=.35\textwidth,viewport= 400 40 5000 3600,clip]{evolution_220_view_paper_hel_t0.pdf}}
\subfigure{
\includegraphics[width=.35\textwidth,viewport= 400 40 5000 3600,clip]{evolution_220_view_paper_hel_t100.pdf}}
\subfigure{
\includegraphics[width=.24\textwidth,viewport= 4 4 392 392,clip]{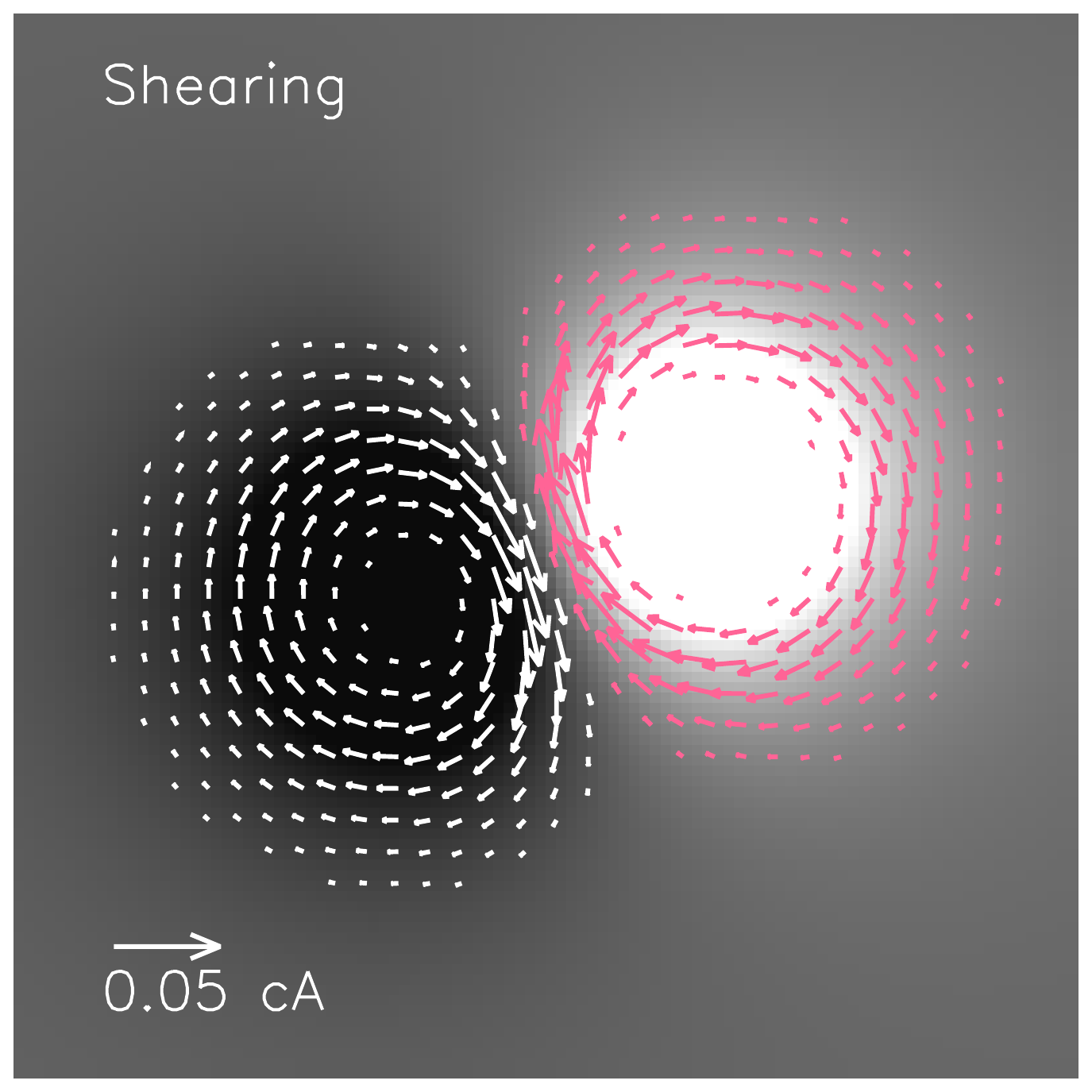}}

\caption{Evolution of the magnetic field for the shearing phase common to all the simulations (left/middle panels) and the applied boundary motions (right panel). The magnetic field lines are color coded with the current density, yellow/red means higher current density.  White/black color indicate positive/negative $B_{z}(z=0)$. White/magenta arrows indicate flows applied to the negative/positive polarity.} %where the system becomes nominally torus unstable, i.e., 
\label{Fig:Shearing}
\end{center}
\end{figure*}
%-----------------------------------------------------------------------------------

\subsection{Initial condition}
\label{Sec:InCon}

The initial condition for the magnetic field, common to all simulations, consists of an asymmetric and bipolar active region generated by two unbalanced sub-photospheric monopoles (see Figure~\ref{Fig:Shearing}, left panel). In the non-dimensional units of the simulation \citep[cf. Section 2.4 of ][for a possible choice of dimensional units]{Aulanier10}, the initial density in the volume is  $\rho(t=0) = B^2(t=0)$, such that the initial Alfv\'{e}n speed is $c_A(t=0)=1$, while the initial velocity field is $\mathbf{u}(t=0)=0$. 

We impose `open' boundary conditions for all the boundaries apart from the boundary at $z=0$, i.e., the photospheric boundary, where line-tied boundary conditions are applied instead \citep{Aulanier05}.  We notice that, as a result of the applied boundary motions and field dynamics, the configuration of the field is naturally expanding and flux is free to leave the simulation box trough lateral and top boundaries throughout the simulations.

\subsection{The common shearing phase}
\label{Sec:Shearing}

For all the simulations, the initial potential magnetic field is evolved into a current-carrying magnetic field by imposing asymmetric vortices centered around the local maxima of $|B_z(z=0)|$. 

Figure~\ref{Fig:Shearing} (right panel) shows the applied flow field. By design these boundary flows induce shear close to the polarity inversion line (PIL) of the active region and  affect the periphery of the active region only mildly. Moreover, the flows are tangent to the iso-contours of $B_z(z=0)$, therefore, during this phase the distribution of $B_z$ at the photospheric boundary $z=0$ remains unchanged. Since the major component of the flow field during this phase consist of shearing motions close to the PIL, we refer to this phase as the shearing phase and to these motions as shearing motions. 

The shearing flows are applied from t$\simeq$10$t_A$ until t$\simeq$100$t_A$. At the end of this phase, the magnetic field configuration is characterized by a highly sheared, current-carrying magnetic arcade surrounded by a quasi-potential background field anchored around the center of the magnetic polarities (see Figure~\ref{Fig:Shearing}, middle panel). 

To ensure that the normal component of the magnetic field at the boundary remains unchanged, during this phase the photospheric diffusion is set to $\eta_{\text{phot}} = 0$.  The coronal diffusion and pseudo-viscosity are set to $\eta_{\text{corona}}=4.8 \times 10^{-4}$ and $\nu' = 25$, respectively \citep[see Section 2.3 of][]{Zuccarello15}.

\subsection{Control case: the non-eruptive run}
\label{Sec:RunB}

As term of comparison throughout this paper we also include a non-eruptive control run obtained by avoiding the photospheric driving phase of Section~\ref{Sec:Driving}. For this run at the photospheric boundary we impose $\mathbf{u}$(t$\gtrsim$100$tA$) = 0 and $\eta_{\text{phot}}=4.8 \times 10^{-4}$ for 100$t_A \lesssim$ t $\lesssim$ 164$t_A$ and $\eta_{\text{phot}}=0$ for t$\gtrsim$164$t_A$. The coronal diffusion and pseudo-viscosity are the same as the shearing phase for 100$t_A \lesssim$ t $\lesssim$ 164$t_A$ and they are increased by a factor  4.37 and 1.67, respectively for t$\gtrsim$164$t_A$. 

The diffusion and pseudo-viscosity parameters have the same time dependence as the other four simulations runs. This allows us to distinguish effects of direct diffusion, which we expect to be similar for identical parameters, from the run-specific dynamic due to different evolution of energy and helicity.

\subsection{The flux-rope formation phases}
\label{Sec:Driving}

From t$\simeq$105$t_A$ the flux rope formation phase starts. During this phase at the line-tied boundary we apply four different types of photospheric motions. 

\begin{figure*}
\begin{center}
\subfigure{
\includegraphics[width=.322\textwidth,viewport= 700 40 5000 3400,clip]{evolution_220_view_paper_hel_t148.pdf}}
\subfigure{
\includegraphics[width=.322\textwidth,viewport= 700 40 5000 3400,clip]{evolution_220_view_paper_hel_t184.pdf}}
\subfigure{
\includegraphics[width=.322\textwidth,viewport= 700 40 5000 3400,clip]{evolution_220_view_paper_hel_t220.pdf}}
\subfigure{\label{RunC-vfield}
\includegraphics[width=.24\textwidth,viewport=4 4 392 392,clip]{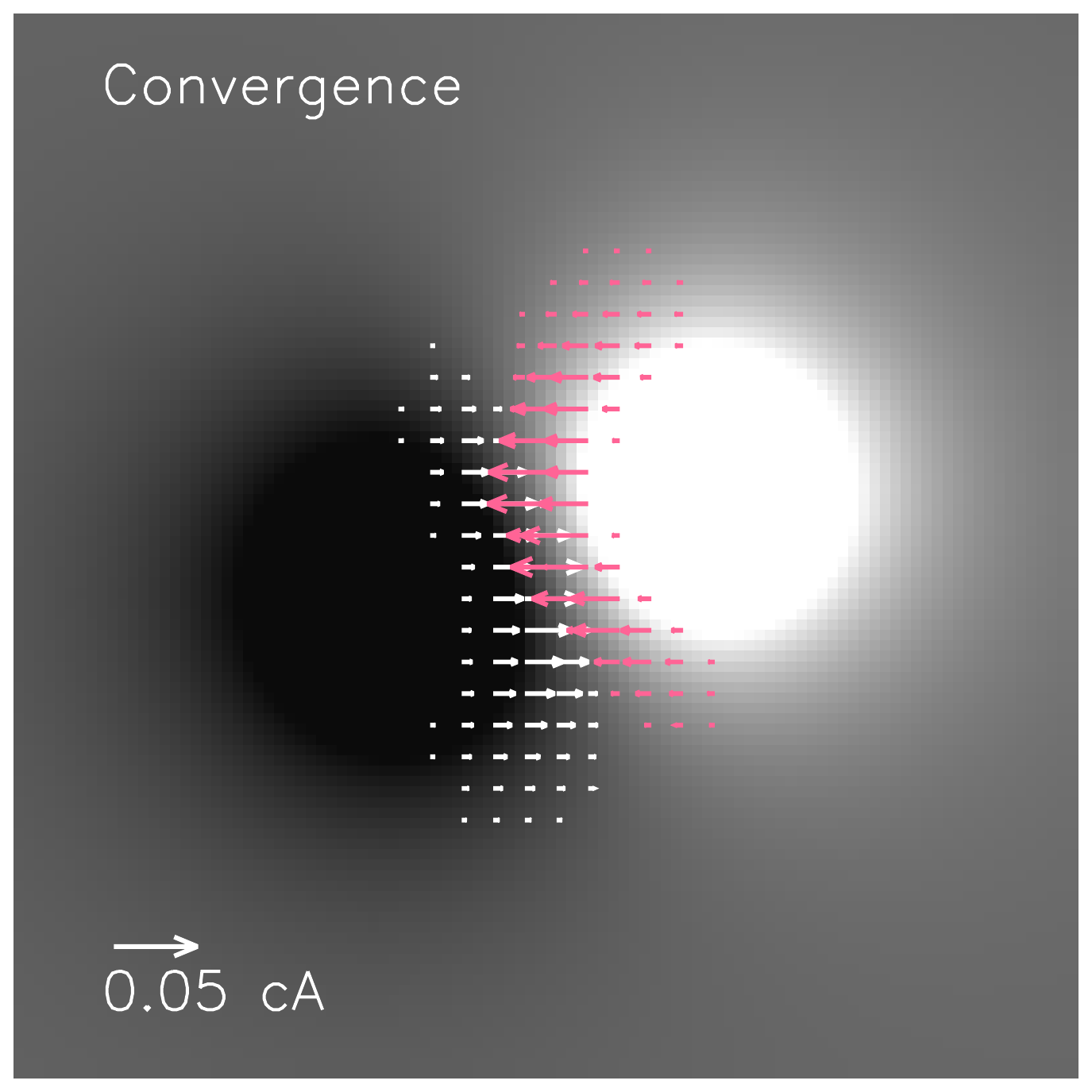}}
\subfigure{\label{RunS-vfield}
\includegraphics[width=.24\textwidth,viewport=4 4 392 392,clip]{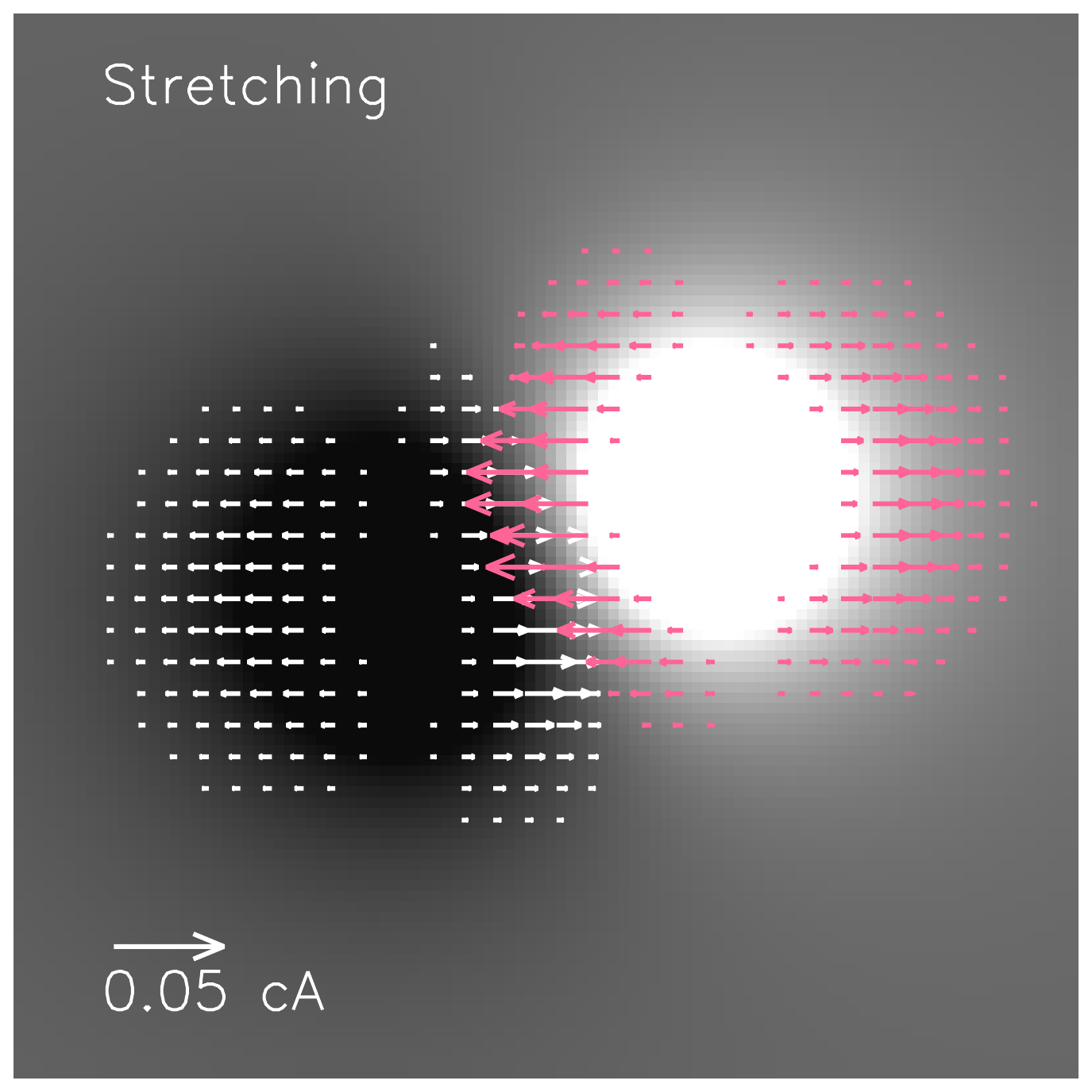}}
\subfigure{\label{RunD1-vfield}
\includegraphics[width=.24\textwidth,viewport=4 4 392 392,clip]{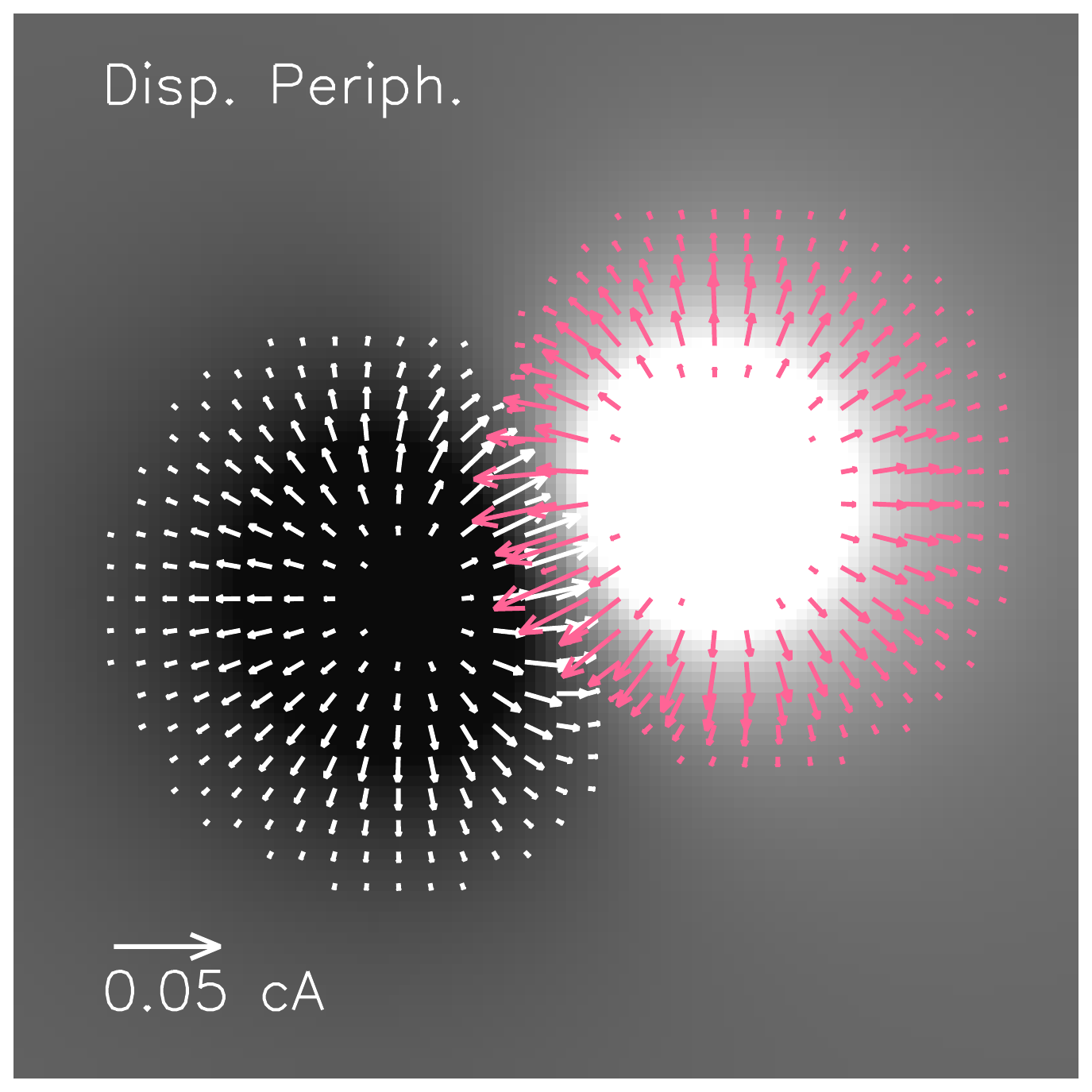}}
\subfigure{\label{RunD2-vfield}
\includegraphics[width=.24\textwidth,viewport=4 4 392 392,clip]{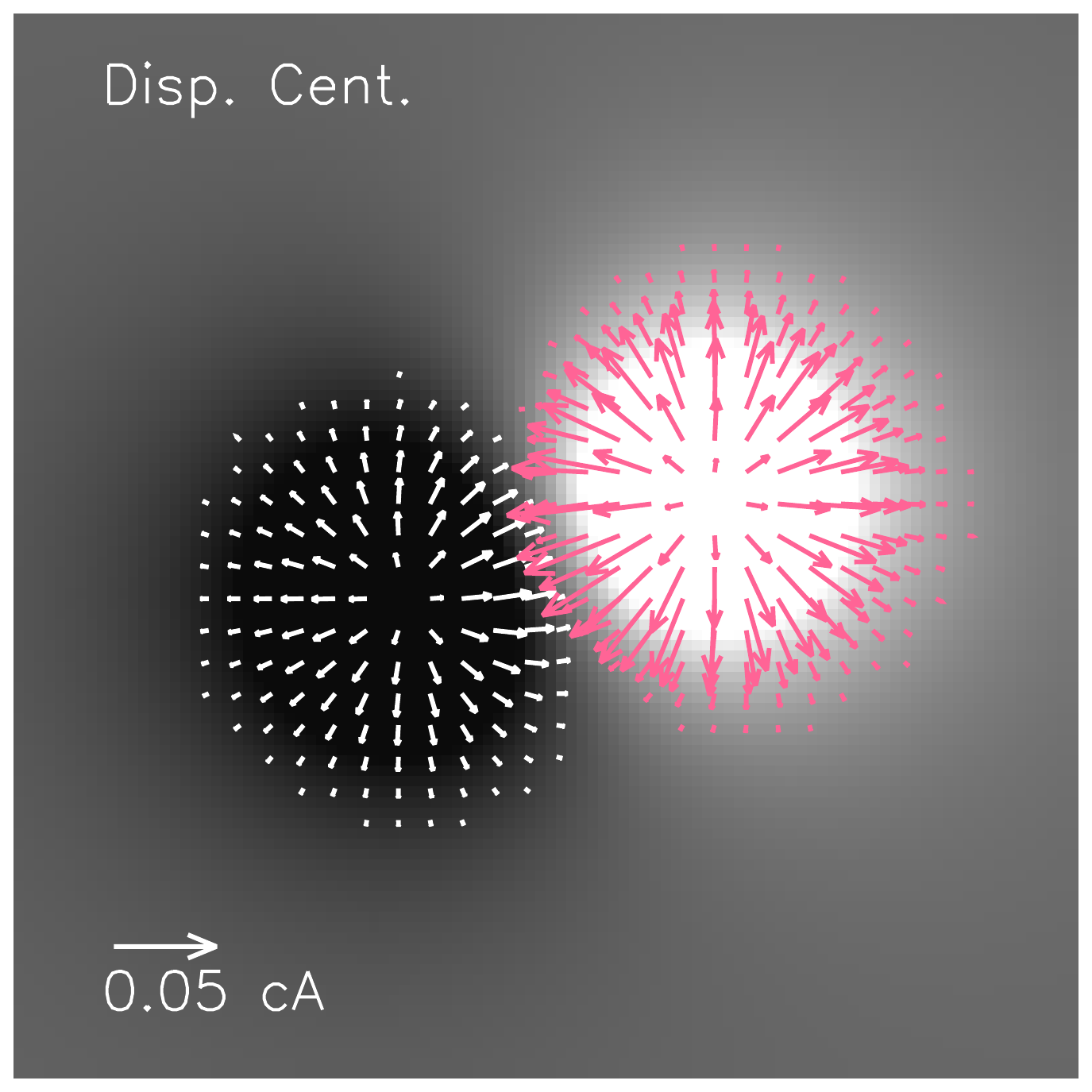}}
\caption{Evolution of the magnetic field for the Dispersion Peripheral run during the flux rope formation phase (top panels) and the applied boundary motions (bottom panels) for the four different simulations runs. Colour scheme is the same as Figure~\ref{Fig:Shearing}.}
\label{Fig:Evolution}
\end{center}
\end{figure*}

Figure~\ref{Fig:Evolution} (bottom panels) shows the applied boundary motions. The four different velocity fields aim to mimic flow patters typically observed on the Sun. The four different velocity fields result in four different simulations runs labeled as Convergence, Stretching, Dispersion Peripheral  (Disp. Periph.) and Dispersion Central (Disp. Cent.). 

The run labeled Convergence  is characterized by flows that only have an horizontal component and are applied only in the proximity of the PIL. These flows result in the advection of photospheric magnetic field towards the PIL, but do not affect the central and peripheral parts of the active region. 

In the run labeled Stretching the flows  are now applied not only in the proximity of the PIL but also in the periphery of the active region. The effect of these flows is to induce an asymmetric stretching of $B_z(z=0)$.     

Finally, the runs labeled as Dispersion Peripheral  and Dispersion Central  are characterized by flows that spreads radially from the center of the magnetic polarities. The  difference between the two flow patterns is the size of the portion of the magnetic polarities that is affected by the flow. In Dispersion Peripheral only the periphery of the magnetic polarities are subjected to the flows resulting in a peripheral dispersion of the magnetic field, while in Dispersion Central a larger region of the polarity is subjected to these flows resulting in a more significant diffusion of the magnetic polarities.

Figure~\ref{Fig:Evolution} (bottom panels) shows that all the flows have a component that advects oppositely directed vertical magnetic field towards the PIL. To allow the cancellation of this oppositely directed magnetic flux, during this phase the photospheric diffusion is set to $\eta_{\text{phot}}=4.8 \times 10^{-4}$. The coronal diffusion and pseudo-viscosity are kept the same as in the shearing phase. 

The response of the solar corona to the applied boundary flows for the  Dispersion Peripheral run  is shown in Figure~\ref{Fig:Evolution} (top panels). As a consequence of the cancellation of magnetic flux around the PIL, a magnetic flux rope is formed through magnetic reconnection at a bald-patch separatrix \citep{Demoulin96}. This reconnection process transfers sheared, arcade-like magnetic flux into the flux rope, eventually increasing the total current within it, and driving its slow rise up to a point when the torus instability sets in and the flux rope undergoes a full eruption. A similar mechanism yields the formation of a flux rope in the other runs as well \citep[see ][for additional details]{Zuccarello15}. The flux rope formation phase ends at the time of the eruption, which happens at a  different time in the four simulations.

%-----------------------------------------------------------------------------------
% Flows Figure
%-----------------------------------------------------------------------------------

%-----------------------------------------------------------------------------------

\subsection{The eruption phase}
\label{Sec:Erupt}

The onset of the torus instability is determined through a series of relaxation runs in which the photospheric boundary flows are gradually re-set to zero using a ramp-down time profile of total time width $\Delta t =6 t_{\text{A}}$. 

In other words, for each of the four experimental set-up, dozens of simulations have been performed in which the applied flows was imposed for different durations before being smoothly stopped. Only when the boundary flow was imposed long enough a marked eruption is observed. If the boundary flows are stopped before the instant $t_{\text{I}}$ the system either relaxes to a new equilibrium or the
flux rope undergoes an extremely deflected eruption \citep[see ][ for more details]{Zuccarello15}. However, if the boundary flows are stopped at (or after) $t_{\text{I}}$, the flux rope undergoes a full eruptions and expands in the numerical domain. Since the four simulations have different photospheric flows evolutions, the exact time at which the instability sets in is different for the four cases. By stopping the photospheric driver at different instants in time and letting the system evolve under the effect of the residual Lorentz force, \cite{Zuccarello15} have shown that the onset of the instability leading to full eruptions occurs at $t_{\text{I}}$ = 196, 214, 220 and  164$t_{\text{A}}$ for the Convergence, Stretching, Dispersion Peripheral and Dispersion Central runs, respectively. It should be noted that the time  $t_{\text{I}}$ corresponds to the middle of the ramp-down time profile, therefore the boundary flows are zero only for $ \bar{t}_{\text{I}} \gtrsim t_{\text{I}} +3 t_{\text{A}}$. The vertical lines in all the figures of the present paper indicate the time $ \bar{t}_{\text{I}}$. 

By using the same ramp-down time profile, at time $t_{\text{I}}$ we also re-set the photospheric diffusion to zero. For numerical stability reasons, at the same time $t_{\text{I}}$ and by using a similar ramp-up time profile we increase the coronal diffusion by a factor 4.37 and the pseudo-viscosity by a factor 1.67. Since we focus on the triggering of the instability, in the following, only the evolution until 10$t_{\text{A}}$ after the time $\bar{t}_{\text{I}}$ of each simulation is shown. However, all simulations were continued for long after that time \citep[see ][]{Zuccarello15}.

\section{Magnetic helicity and energy decompositions }
\label{Sec:Helicity}

The  magnetic helicity  $H$ of a magnetic field $\vB$ in a volume $\vol$ is defined as:
\BE
\label{eq: H_original}
H = \int_{\vol} \vA \cdot \vB \dV \ ,
\EE
where $\vA= \nabla \times \vB$ is the vector potential. This quantity is gauge invariant only when the magnetic field $\vB$ is fully contained inside the volume $\vol$, e.~g., when the magnetic field is tangential to the surface $\surf$ that bounds $\vol$. This condition is rarely satisfied in the magnetic field systems that are of interest in solar physics, i.e., open coronal volumes. 

Following the work of \cite{Berger84}, \cite{Finn85} showed that in the case where  $\vB$ is not fully contained in $\vol$ a quantity that is gauge invariant by definition and it is  better suited to characterize the system is the relative magnetic helicity: 
\BE %____________________________________ 
\label{eq:Hv}
\Hv = \int_{\vol} (\vA+\vAp ) \cdot (\vB-\vBp ) \dV \ ,
\EE
with  $\vAp$ the vector potential of the potential field $\vBp=\curlAp$ that has the same distribution of the normal component of  $\vB$ on the bounding surface.  

A possible decomposition of Equation~\ref{eq:Hv} is \citep{Berger03}:
 \BA %____________________________________ 
\Hv  &=& \Hj + \Hpj \ , \quad \text{with} 
               \label{eq:HDecomp}\\
\Hj   &=& \int_{\vol} (\vA - \vAp)  \cdot (\vB-\vBp) \dV \ ,
               \label{eq:Hj}\\
\Hpj &=& 2\int_{\vol} \vAp \cdot (\vB-\vBp) \dV \ ,
               \label{eq:Hpj}
\EA
where $\Hj$ is the  magnetic helicity of the non-potential, or current carrying, component of the magnetic field, $\vBj=\vB-\vBp$, and $\Hpj$  is a volume threading term involving both $\vBp$ and $\vBj$. %The field $\vBj$ is contained within the volume $\vol$ so it is also called the closed field part of $\vB$. 
Because  $\vB$ and $\vBp$ have the same normal distribution on $\surf$, both $\Hv$, $\Hj$ and $\Hpj$ are separately gauge invariant. Similarly to \citet[][cf. Section 4.1]{Pariat17}, in the present paper, the quantities $\vBp$, $\vAp$, and $\vA$ are computed using the method of \cite{Valori12}.  

The different flux rope formation phases are associated with different photospheric boundary motions that result in different evolutions of  $B_z(z=0)$. As a result, the magnetic flux is different for the different simulations. In order to account for these differences when comparing the various helicity decompositions at the moment of the eruption we consider their normalized value, i.e.,  $\Hv(t)/\Phi^2(t)$, $\Hj(t)/\Phi^2(t)$ and $\Hpj(t)/\Phi^2(t)$, where $\Phi(t)=\frac{1}{2}\int_{z=0} \left | B_z(z=0,t) \right | \, d\rm{x} d\rm{y}$. 

In the present paper, the different decompositions of the magnetic energy are computed following the approach discussed in \citet{Valori13}, where the magnetic energy of a magnetic field with finite non-solenoidality ($\nabla\cdot\mathbf{B}\ne0$), can be decomposed as:  
\BE %____________________________________
\label{eq:thomson}
\E = \Ep +\Ej +\Ens \ ,
\EE
where $\Ep$ and $\Ej$  are the energies associated with the potential and current-carrying solenoidal contributions,  and $\Ens$,  is the sum of the artifact non-solenoidal contributions \citep[see Eqs.~(7,8) in][ for the corresponding expressions]{Valori13}.  For purely solenoidal fields $E_{ns}$ is  zero, however, finite non-solenoidality is generally present when discrete numerical meshes are considered. As discussed in \citet{Valori16} the non-solenoidality of the field  actually affects the precision of the helicity computations. For the simulations presented here the average non-solenoidality is $\Ens / \E \simeq 0.02$. We note that in order to apply the method of \citet{Valori12}, the non-uniform grid used to perform the simulations has been interpolated into a uniform grid and the divergence values are increased by the interpolation. While this values are not representative of the quality of the simulations themselves, they nevertheless allow us to estimate the precision of the magnetic helicity computations discussed here. According to the results of \citet{Valori16} the precision of our helicity computations is  $\lesssim$2\%.

\section{Evolution of the magnetic flux}
\label{Sec:MagFlux}

\begin{figure}%[!t]
\centering
\includegraphics[width=.42\textwidth,viewport= 10 18 560 548,clip] {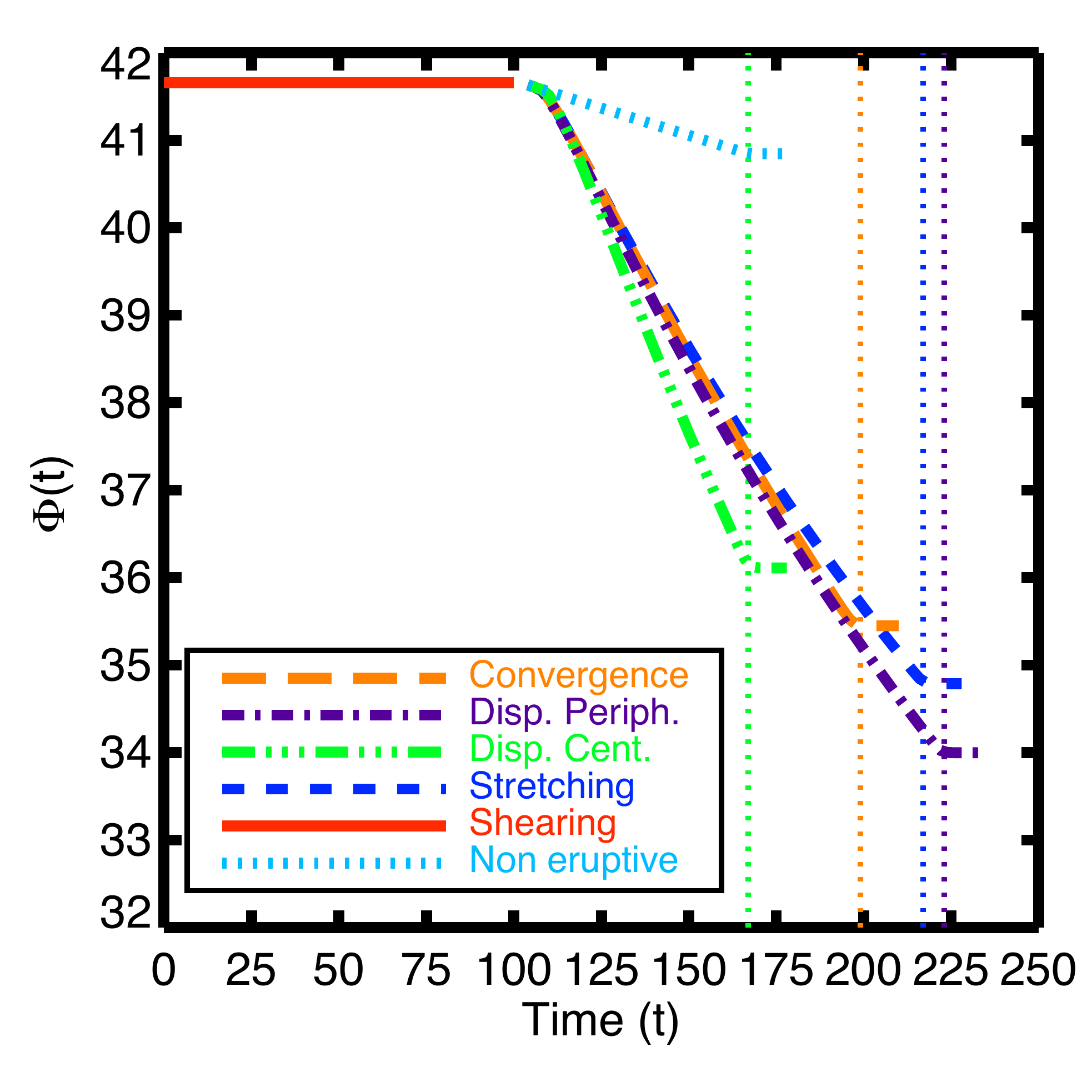}
\caption{Time evolution of the photospheric magnetic flux for the common shearing phase (red), for the non eruptive run (cyan) and for the four eruptive runs.  
\label{Fig:Flux}}
\end{figure}

The evolution of the photopsheric magnetic flux as a function of time and for all the simulation runs is presented in Figure~\ref{Fig:Flux}. 
The magnetic flux is constant during the common shearing phase. This is a consequence of the design of the boundary motions, which do not change the phostospheric distribution of $B_z$, and of the  fact that the photospheric diffusion is $\eta_{\text{phot}}=0$ during this phase. 

In the control non-eruptive run, where the flows are set to zero and only a finite photospheric diffusion is imposed at the boundary, only $\lesssim$2\% of the initial photospheric flux is diffused within $\sim$60 $t_{\text{A}}$.

During the flux rope formation phase, opposite magnetic flux is advected towards the PIL in all four simulations. Combined with a finite photopsheric diffusion, this results in the cancellation of about 13-18\%  of the initial photospheric flux at the moment of the onset of the eruption. 

From $\bar{t}_{\text{I}}$ both photospheric flows and diffusion are re-set to zero and the photospheric flux remains constant until the end of the simulation. 

The change in the slope of the photospheric flux that we observe towards the end of the flux rope formation phases in Figure~\ref{Fig:Flux} is essentially due to the change in the forcing of the bottom boundary of the simulation, as expected since this quantity is only measured at this boundary, and does not allow to discern the moment of the onset of the instability. 

Finally, we note that at the moment of the eruption the different runs  have reached different values of the magnetic flux.

%-----------------------------------------------------------------------------------
% Magnetic flux Figure
%-----------------------------------------------------------------------------------

%-----------------------------------------------------------------------------------
%-----------------------------------------------------------------------------------
% Full time range Figure
%-----------------------------------------------------------------------------------

\begin{figure*}%[!t]
 \subfigure{	
 \includegraphics[width=.32\textwidth,viewport= 0 18 558 558,clip] {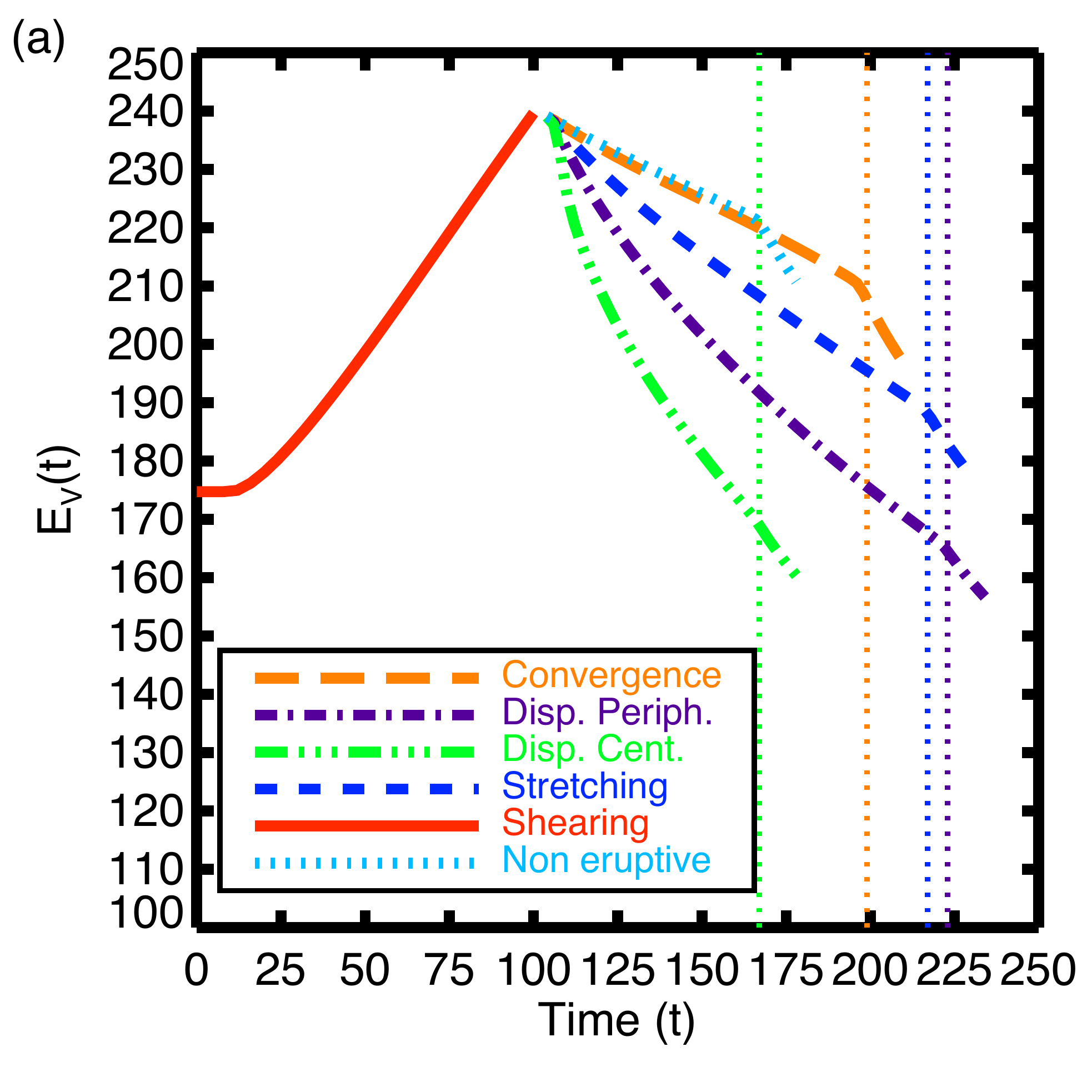}\label{Fig:E}
 }
 \subfigure{	
 \includegraphics[width=.32\textwidth,viewport= 0 18 558 558,clip] {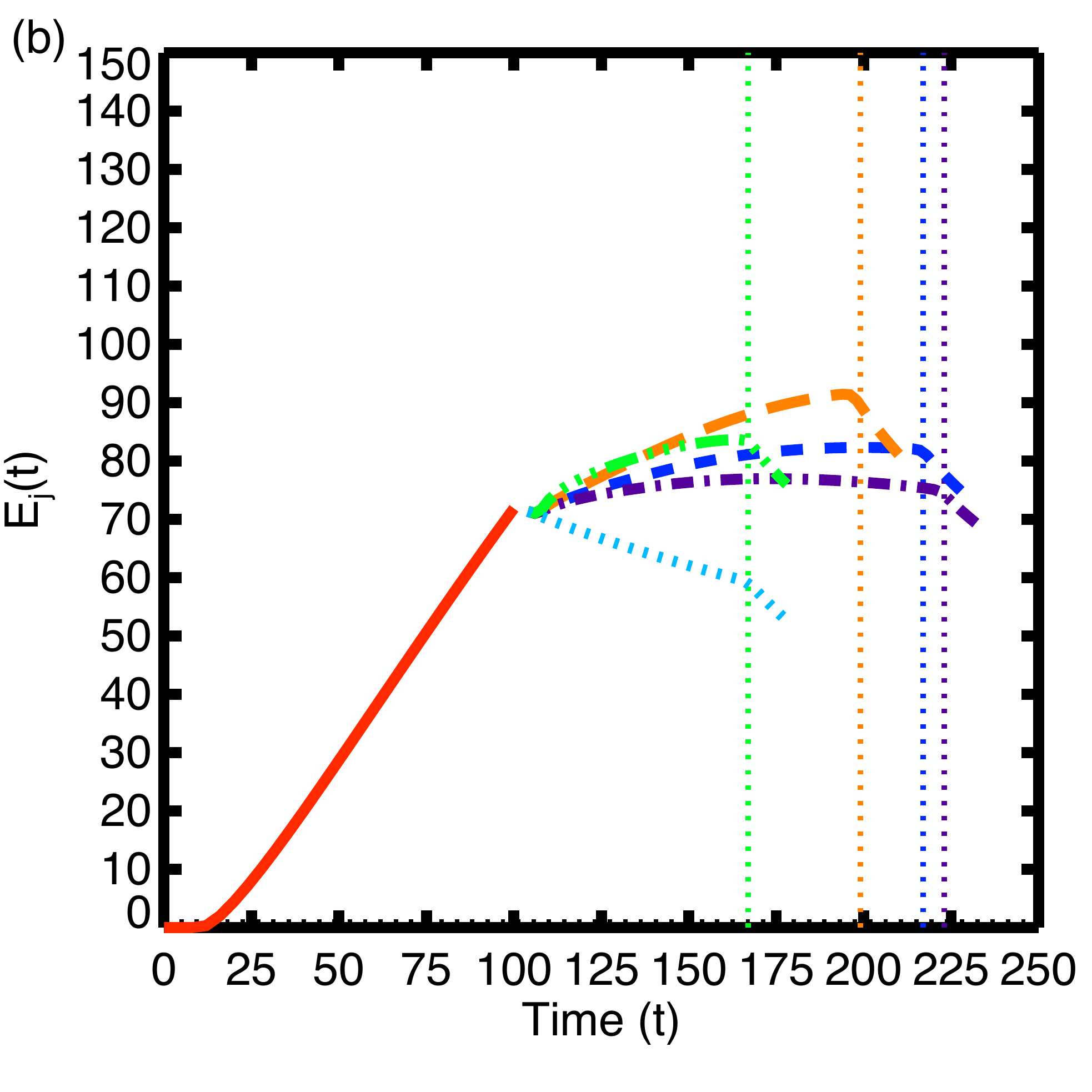}\label{Fig:Efree}
 }
\subfigure{	
 \includegraphics[width=.32\textwidth,viewport= 0 18 558 558,clip] {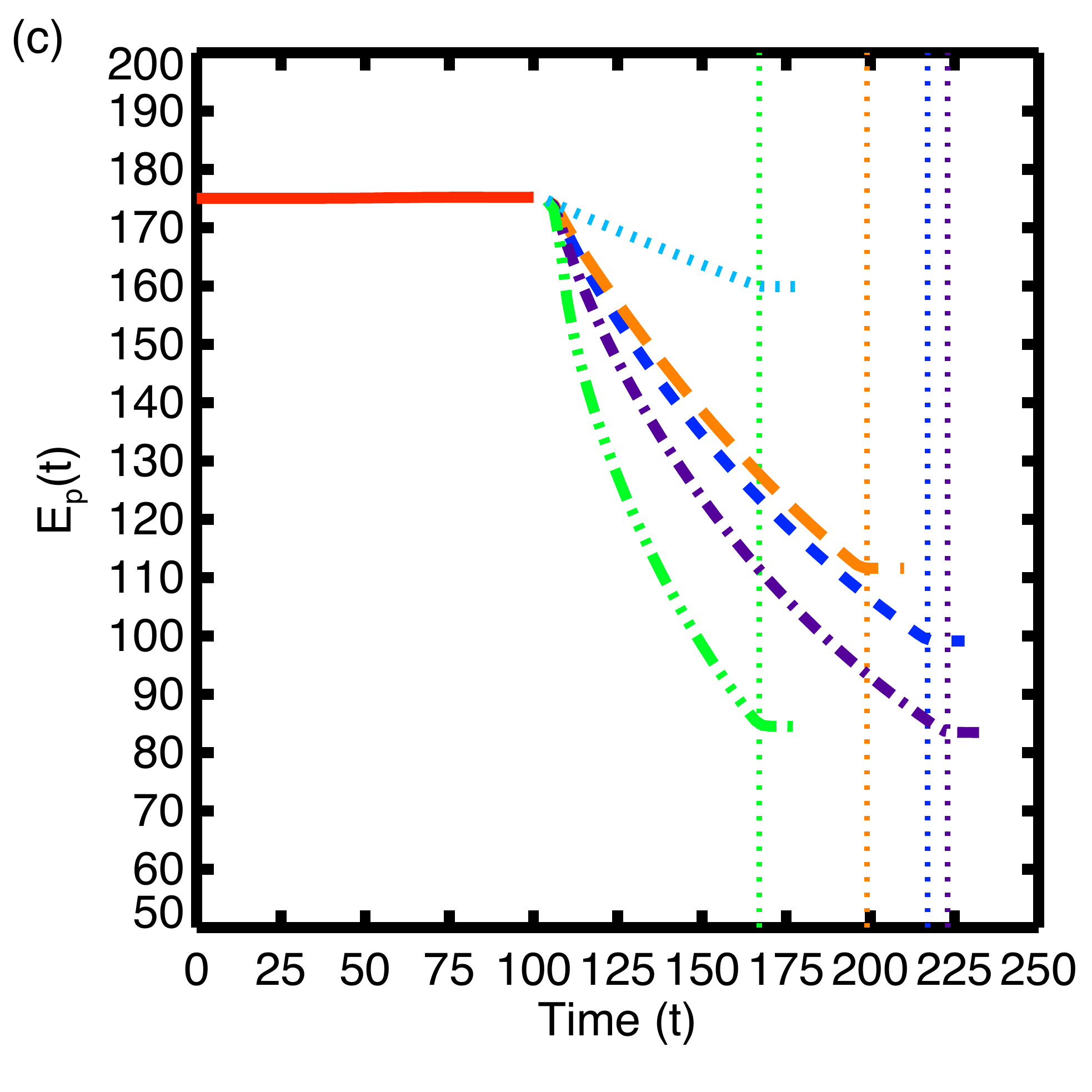}\label{Fig:Epot}
 }
\subfigure{	
 \includegraphics[width=.323\textwidth,viewport= -12 18 557 558,clip] {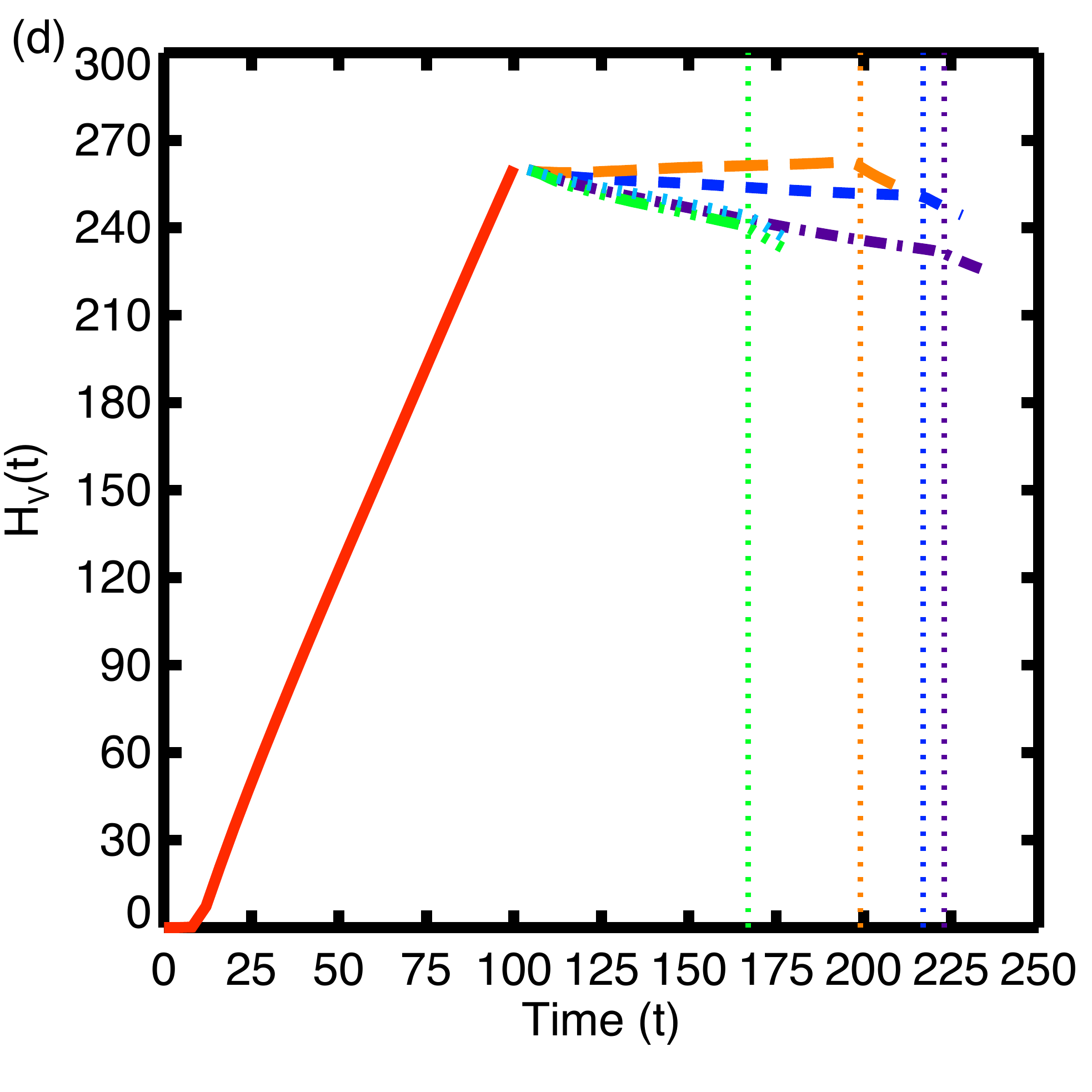}\label{Fig:Hv}
 }
 \subfigure{	
 \includegraphics[width=.323\textwidth,viewport= -8 18 559 558,clip] {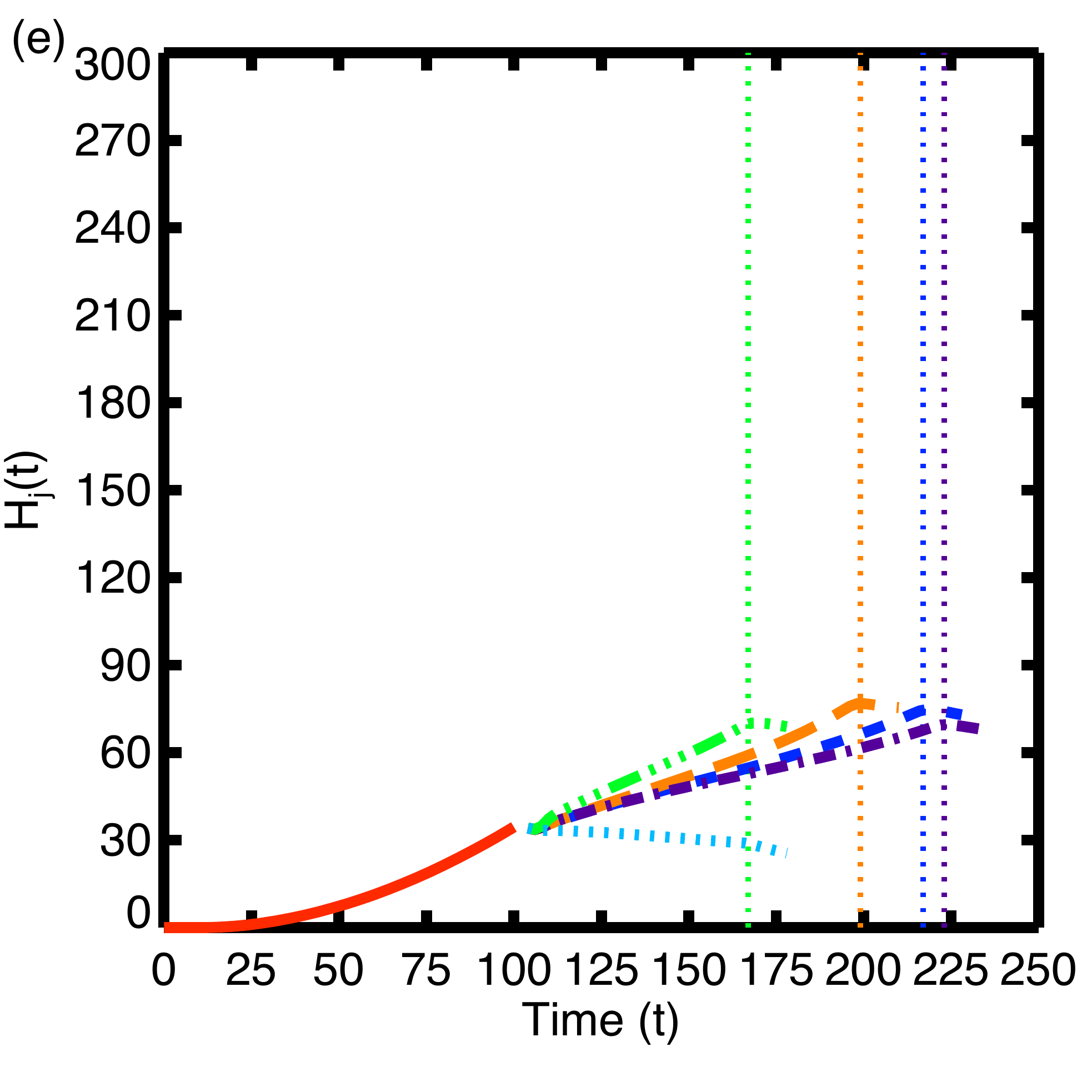}\label{Fig:Hj}
 }
 \subfigure{	
 \includegraphics[width=.33\textwidth,viewport= -24 18 558 558,clip] {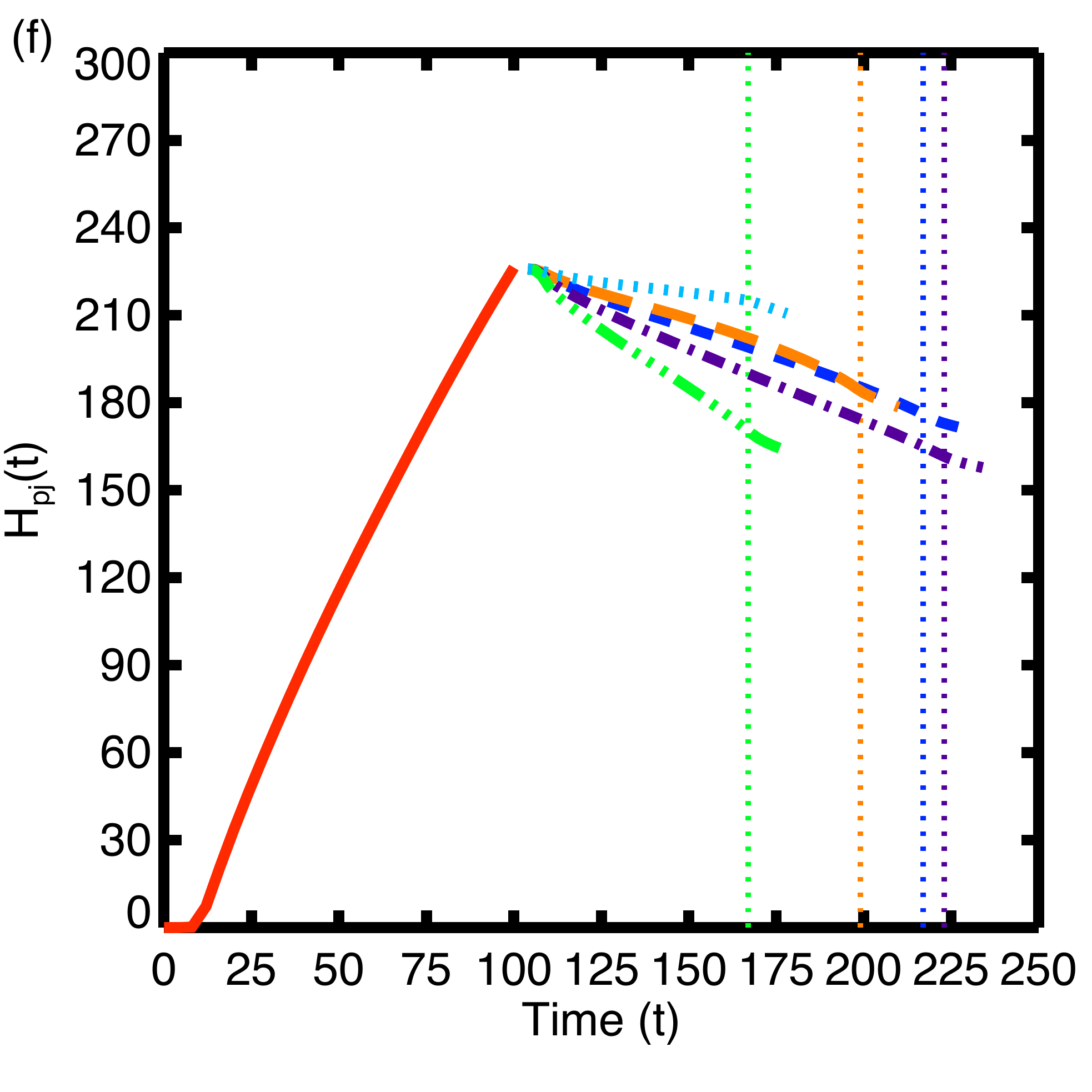}\label{Fig:Hpj}
 }
\caption{Time evolution of the different energy (top) and helicity (bottom) terms for the  shearing and flux rope formation  phases.
\label{Fig:H}}
\end{figure*}

%-----------------------------------------------------------------------------------

%-----------------------------------------------------------------------------------
% Shifted time range Figure
%-----------------------------------------------------------------------------------

\begin{figure*}%[!t]
 \subfigure{	
 \includegraphics[width=.32\textwidth,viewport= 0 18 558 558,clip] {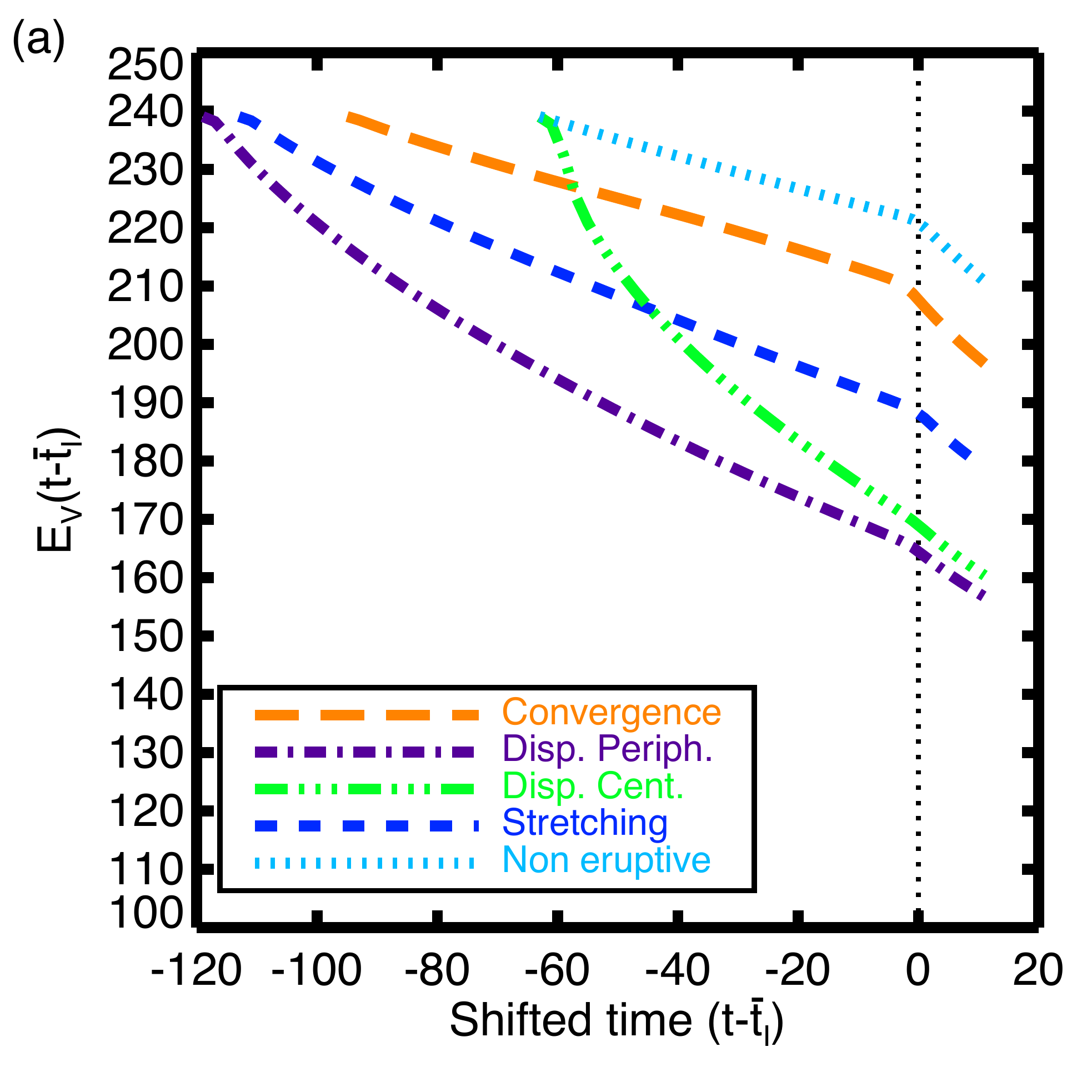}\label{Fig:E_shift}
 }
 \subfigure{	
 \includegraphics[width=.32\textwidth,viewport= 0 18 558 558,clip] {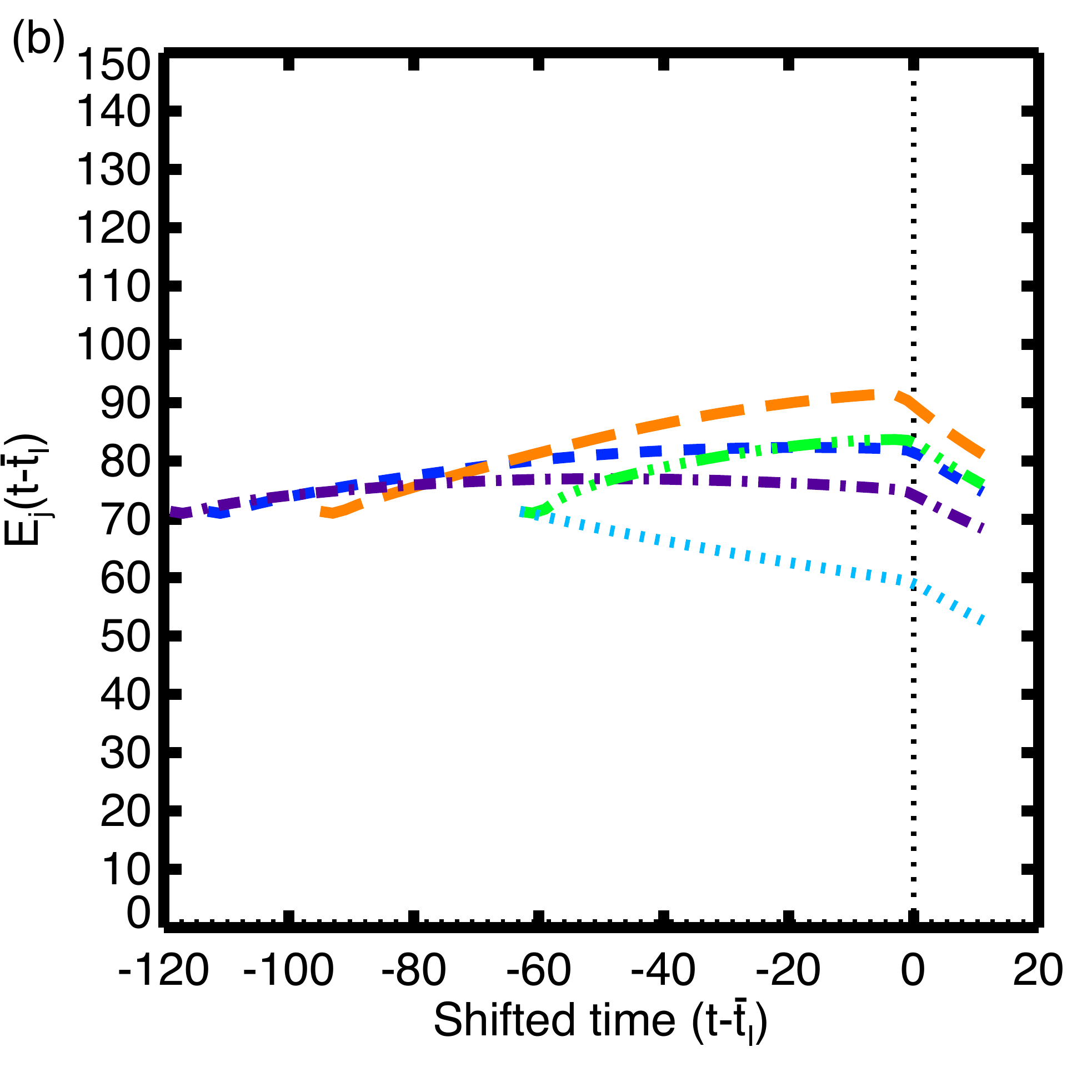}\label{Fig:Efree_shift}
 }
 \subfigure{	
 \includegraphics[width=.32\textwidth,viewport= 0 18 558 558,clip] {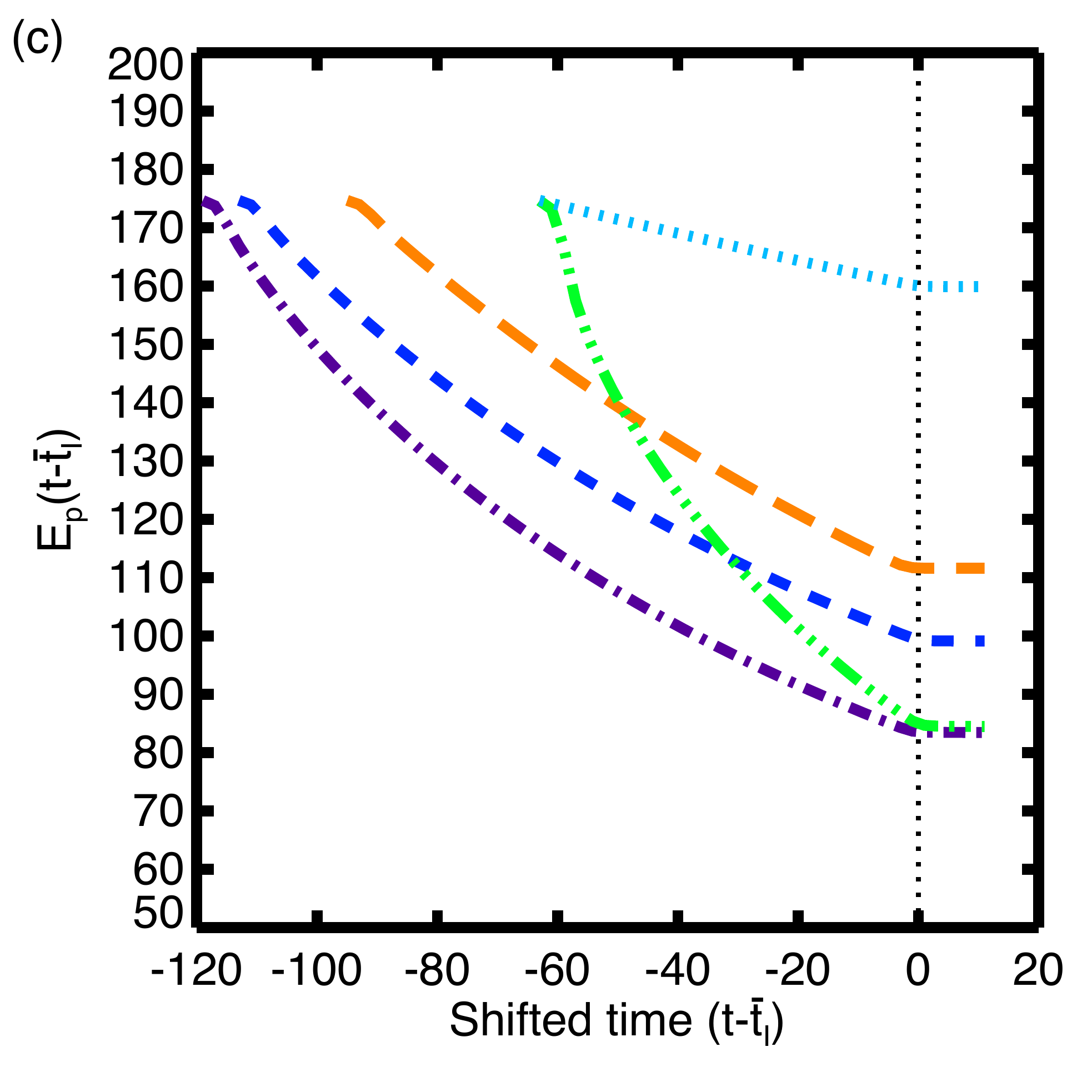}\label{Fig:Epot_shift}
 }
 \subfigure{	
 \includegraphics[width=.32\textwidth,viewport= 4 18 557 558,clip] {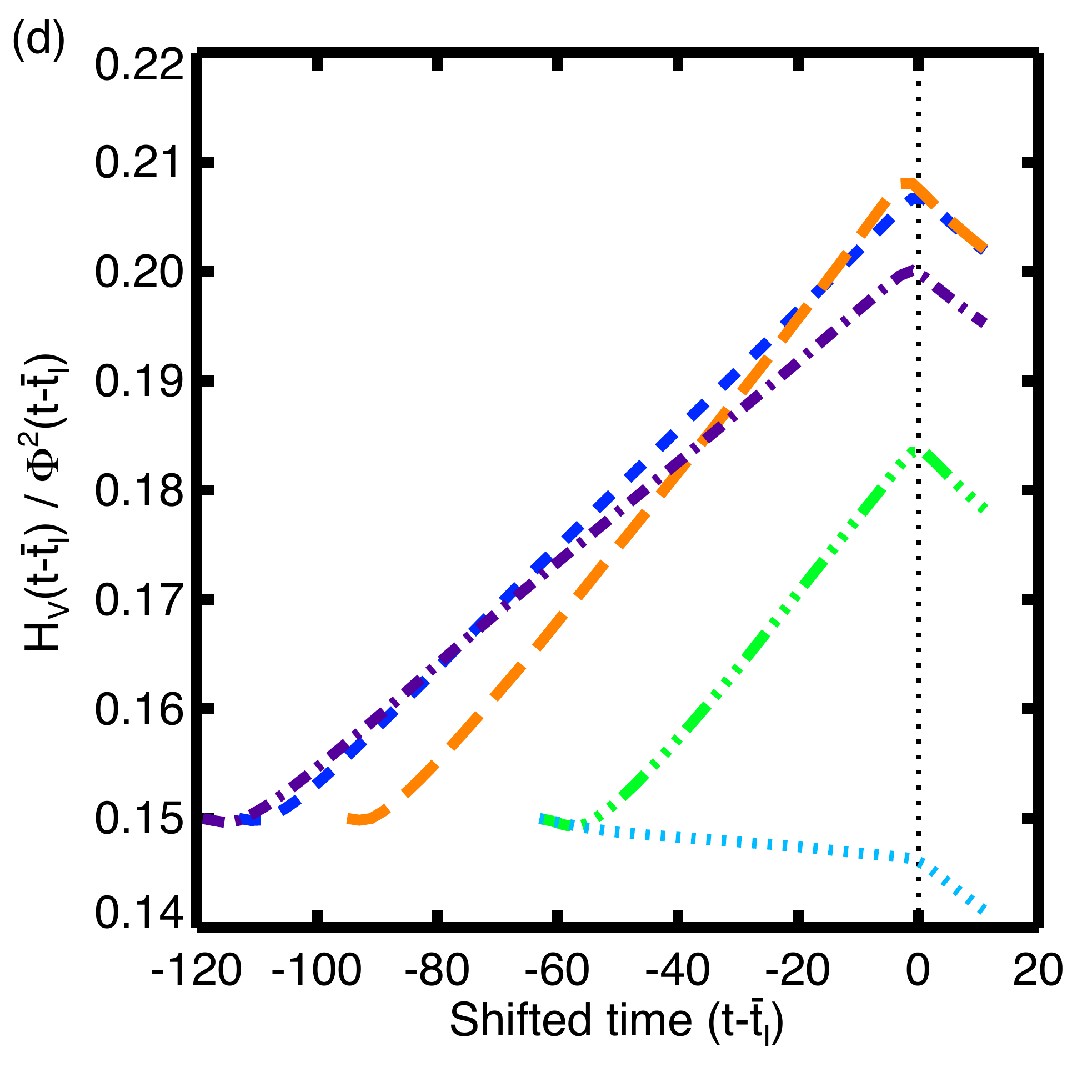}\label{Fig:Hv_shift}
 }
 \subfigure{	
 \includegraphics[width=.32\textwidth,viewport= 7.5 18 554 558,clip] {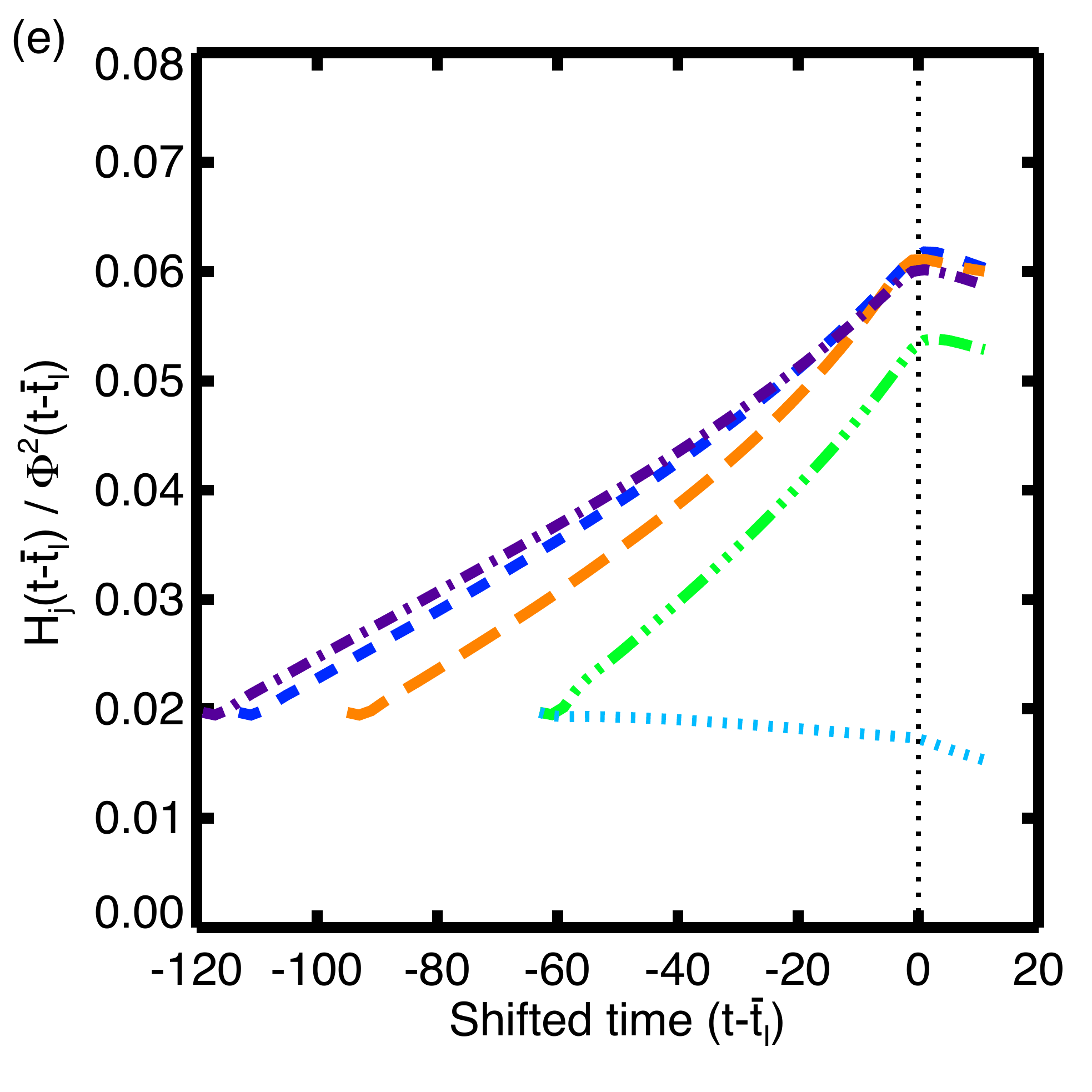}\label{Fig:Hj_shift}
 }
 \subfigure{	
 \includegraphics[width=.328\textwidth,viewport= -10 18 554 558,clip] {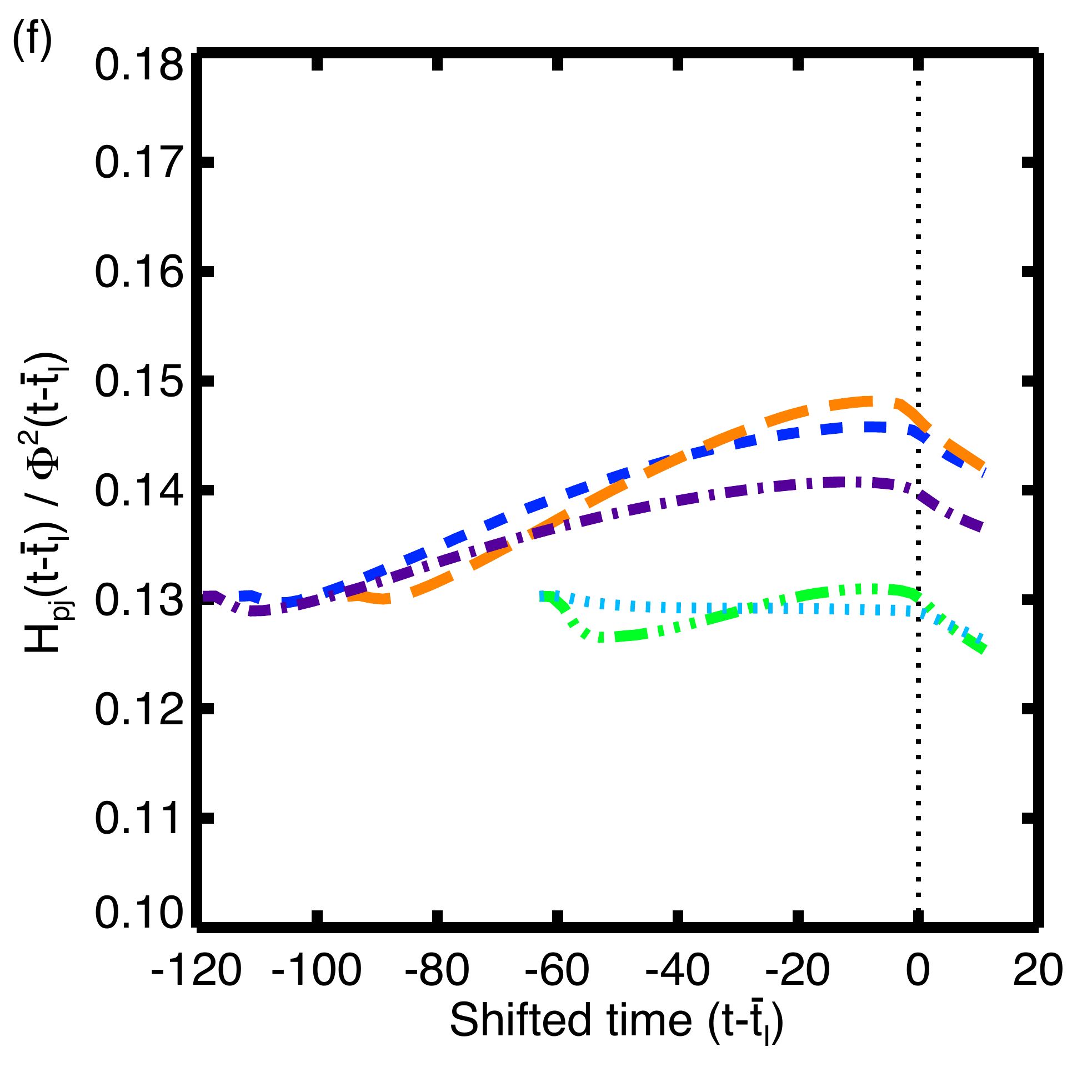}\label{Fig:Hpj_shift}
 }
\caption{Time evolution of the different energy (top) and normalized helicity (bottom) terms during the flux rope formation phase. The time scale is shifted so that the time are given with respect to $\bar{t}_I$.
\label{Fig:H_shift}}
\end{figure*}
%-----------------------------------------------------------------------------------

\section{Trends in magnetic energy and helicity }
\label{Sec:Trends}

In this Section we discuss the trends in the time evolution of the magnetic energy and of the magnetic helicity for the different runs.

\subsection{Comparison between the shearing and flux rope formation phases}
\label{Sec:ShearDrivPhas}

Figure~\ref{Fig:H}(top panels) shows the evolution of the different energy decompositions. During the common shearing phase, i.e., from t$\simeq$10$t_{\text{A}}$ to t$\simeq$100$t_{\text{A}}$, $\E$  shows a linear increase up to about 37\% of its initial value. A comparison between $\Ej$ and $\Ep$ (Figures~\ref{Fig:Efree} and \ref{Fig:Epot}) shows that the increase in $\E$ is  due to the increase of $\Ej$. This is expected, since the shearing flows are designed in such a way to not change the boundary distribution of $B_z$, and, hence, of $\Ep$.

In the control non-eruptive run, from t$\simeq$105$t_{\text{A}}$ onward, all the different energy decompositions  display a decrease due to the finite photospheric and coronal diffusion. After t$\simeq$164$t_{\text{A}}$ $\E$ and $\Ej$  continue to decrease even faster (the coronal diffusion is further increased during this phase, see Section ~\ref{Sec:RunB}), while $\Ep$ is now constant, since $\eta_{phot} $ is re-set to zero during this phase.  

During the different flux rope formation phases, i.e., from t$\simeq$105$t_{\text{A}}$ to t$\simeq t_{\text{I}}$, $\E$ decreases, and this is the case up to the end of the simulations (not shown in the Figure). Despite the total magnetic energy decreases, Figure~\ref{Fig:Efree} shows that, apart from the control case,  $\Ej$ actually increases during the flux rope formation phase, up to the time t$\simeq t_{\text{I}}$, where it reaches a maximum and starts to decrease. Figure~\ref{Fig:Epot} shows that $\Ep$ decreases during the flux rope formation phase, suggesting that the major reason of the energy decrease during the flux rope formation phase is due to a decrease  of  $\Ep$.  

A comparison between the shearing and the flux rope formation phases shows that the major injection of  $\Ej$ occurs during the shearing phase; the rate of increase of $\Ej$  during the shearing phase is between $\sim$4.5 and $\sim$54 times higher than its rate of increase during the flux rope formation phase. 

The time evolution of $\Hv$ is shown in Figure~\ref{Fig:Hv}. During the common shearing phase $\Hv$ steadily and linearly increases with time. This trend changes during the flux rope formation phase, when the total helicity is either roughly constant or slightly decreases. This is true until the end of  simulations (not shown in the Figure). The situation is different when only the current carrying component of the helicity, i.e., $\Hj$, is considered (Figure~\ref{Fig:Hj}). Similarly to $\Hv$, also $\Hj$ increases during the shearing phase, but with a profile that is somehow different. The linear increasing phase starts with a  delay of about 10-15 $t_{\text{A}}$ with respect to $\Hv$.  

Figure~\ref{Fig:Hj} shows that from t$\simeq$105$t_{\text{A}}$ onward, for the control run, $\Hj$ is either constant or decreases. The situation is different for the other runs, where during the flux rope formation phase $\Hj$  continues to increase  up to $\bar{t}_{\text{I}}$. Figure \ref{Fig:Hj} also shows that $\Hj$  increases at a comparable  rate both during the shearing and the flux rope formation phases. 

The evolution of $\Hpj$ is shown in Figure~\ref{Fig:Hpj}. Similarly to $\Hv$,  during the shearing phase $\Hpj$ steadily increases in time accounting for the major part of the  helicity injection during this phase. From t$\simeq$105$t_{\text{A}}$ onward,  $\Hpj$ for the control run is either constant or decreases.  While the other runs also show the same trend their respective decrease of $\Hpj$ is more significant than the one observed for the control run. Therefore, the decrease of $\Hpj$ during the flux rope formation phase is not only due to the finite diffusion, but likely to a re-distribution of the relative helicity between its different component $\Hj$ and $\Hpj$ (cf. Linan et al. 2018, in prep.).

To summarize the analysis shows that: (1) the largest injection of total magnetic energy and  relative magnetic helicity occurs during the shearing phase; (2) during the shearing phase,  $\Ej$ increases as it is the case for $\Hpj$, i.e., magnetic energy and helicity behave differently during this phase;  (3) at the end of shearing phase,  $\E$ is dominated by  $\Ep$ and $\Hv$  is dominated by $\Hpj$, i.e., magnetic energy and helicity are similar in this aspect; (4) in the flux rope formation phase, both  $\E$ and $\Hv$ decrease, both  $\Ep$ and $\Hpj$ decrease and both $\Ej$ and $\Hj$  increase, i.e., they behave similarly, unlike during the shearing phase; (5)  $\Hj$ is injected with roughly the same rate during the shearing and flux rope formation phases, while this is not the case for $\Ej$  where the most of the  injection occurs during the shearing phase; (6) overall, the flux rope formation phase has helped to strengthen the non potentiality of the field and its relative portion in both $\E$ and $\Hv$ budgets; at the end of the flux rope formation phase, both $\Ej$/$\E$ and $\Hj$/$\Hv$ have increased compared to their value at the start of this phase (see Section~\ref{Sec:Thresholds}).

\subsection{Role of the different boundary flows during the flux rope formation phase}
\label{Sec:DrivPhas}

In this Section we compare the evolution during the different flux rope formation phases, focusing on the similarities and differences between them. 

Figure~\ref{Fig:H_shift} (top panels) shows the evolution of the different energy decompositions during the flux rope formation phases. In order to facilitate the comparison, the curves in Figures~\ref{Fig:H_shift} and \ref{Fig:Ratios} are shifted such as to align the eruption times. The total magnetic energy, $\E$, decreases for all the runs including the non-eruptive one (see Figure~\ref{Fig:E_shift}). However, the Dispersion Central run, which is the run where the major part of the active region is subjected to the convergence flows, displays the fastest decrease of $\E$, while the Convergence run, where only the portion of the active region closets to PIL is subjected to the convergence motions, shows the slowest decrease of $\E$. This latter is actually comparable to the decrease of $\E$ for the control run, where no flows are applied and the energy dissipation is only due to the coronal and photospheric diffusion. 

The time evolution of $\Ej$ is shown in Figure~\ref{Fig:Efree_shift}. A clearer distinction in the trends is visible between the eruptive runs and the non-eruptive one.  For the eruptive simulations, $\Ej$ increases up to the moment of the eruption, while it is always decreasing for the non-eruptive run.  Differently from $\E$, $\Ej$ for the Convergence and Dispersion Central runs follows a very similar trend, despite the fact that these two runs are the ones with the most different flows. Figure~\ref{Fig:Efree_shift} shows that $\Ej$ starts to decrease after the onset of the eruption when driving flow is terminated and coronal dissipation is increased four-folds. A similar initial decrease is also observed for the control run as soon as the system is let to relax under the effect of the increased coronal diffusion.

Figure~\ref{Fig:Epot_shift} shows that, during the flux rope formation phase,  $\Ep$ decreases. Furthermore, the different curves are ordered in the same way as the ones of $\E$, confirming that the major decrease of magnetic energy during the flux rope formation phase is due to the decrease of the energy associated with the potential magnetic field.

%-----------------------------------------------------------------------------------
% Table 1
%-----------------------------------------------------------------------------------

\begin{table*}[!t]
\begin{center}  

 \caption{Values of the different energy and helicity terms at the moment of the onset of the instability.}              
 \label{table:2}      
 \begin{threeparttable}                       
 \begin{normalsize}
 \begin{tabular}{ c | c | c | c | c | c | c | c | c | c | c }        % centered columns (4 columns)
 \hline\hline                 % inserts double horizontal lines
\rule[-2mm]{0mm}{5mm}
 Run                           & $t_{\text{I}} (t_A)$    & $\Phi$    & $\E$  &  $\Ep$               & $\Ej$      & $\Ej/ \E $        & $\Hv$      & $\Hj$     & $\Hpj$     & $\Hj / \Hv$     \\  
\hline  \hline                     
  Convergence                  & 196  	                & 35        & 206.5   &  114.7$\pm3.1$      & 91.8$\pm3.1$   & 0.444$\pm0.015$   & 260$\pm5$  & 77$\pm5$  & 183$\pm5$  & 0.296$\pm0.012$ \\ 
\hline            
 Stretching                    & 214  	                & 35        & 187.4   &  102.8$\pm3.6$      & 84.6$\pm3.6$   & 0.451$\pm0.019$   & 250$\pm5$  & 75$\pm5$  & 175$\pm5$  & 0.300$\pm0.012$ \\ 
\hline           
  Dispersion Peripheral        & 220  	                & 34        & 163.7   &  86.8$\pm3.3$       & 76.9$\pm3.3$   & 0.470$\pm0.020$   & 231$\pm5$  & 70$\pm5$  & 161$\pm5$  & 0.303$\pm0.012$ \\ 
\hline
  Dispersion Central           & 164  	                & 36        & 168.1   &  85.1$\pm0.4$       & 83.1$\pm0.4$   & 0.494$\pm0.002$   & 239$\pm5$  & 70$\pm5$  & 169$\pm5$  & 0.292$\pm0.012$ \\
 \hline\hline                                  
 \end{tabular}
\end{normalsize}
  \begin{tablenotes}
      \small
      \item Note. The value of the different quantities are given at $ \bar{t}_{\text{I}} \gtrsim t_{\text{I}} +3 t_{\text{A}}$, i.e., after the boundary motions are reset to zero. 
    \end{tablenotes}
  \end{threeparttable}
  \end{center}  
\end{table*}

%-----------------------------------------------------------------------------------

The time evolution of the different normalized helicity decompositions is presented in Figure~\ref{Fig:H_shift} (bottom panels). Globally, $\Hv/\Phi^2$ and $\Hj/\Phi^2$ show similar trends and clearly allow to distinguish between the eruptive and non eruptive runs; the two quantities increase for the eruptive runs, while they are roughly constant (although decreasing, largely because of the finite coronal diffusion) for the non eruptive run. This is true until t$\simeq t_I$ when  $\Hv/\Phi^2$ and $\Hj/\Phi^2$ start to decrease for all the runs (including the control run) as a consequence of the increased coronal diffusion.  

A closer look at Figure~\ref{Fig:Hj_shift} shows that, while the different curves follow a very similar trend, some differences exist. An interesting result can be found by comparing the Stretching and Convergence runs. For these two runs the same photospheric motions profile is applied close to the PIL, and the difference only involves the periphery of the active region  (see Figure~\ref{Fig:Evolution}, bottom panels). As a result, sheared arcade flux is advected towards the PIL and eventually converted into flux rope's flux, in a similar fashion for the two runs. The only difference is at the periphery of the active region where part of the overlying magnetic flux is anchored.  This seems to suggest that even the evolution of $\Hj$, which is in principle only related to the current carrying part of the magnetic field, seems to be affected by the evolution of the background field.  This is an example of the non local character of the  magnetic helicity. This result is also consistent with the analysis of the time evolution of $\Hj$ and $\Hpj$ of Linan et al. 2018 (in prep.), which indicates that $\Hj$ is usually not evolving because of boundary flux but is rather transformed from $\Hpj$.

Finally, Figure~\ref{Fig:Hpj_shift} shows that $\Hpj/\Phi^2$ initial decreases during the first stages of the convergence phase and then steadily increases up to few Alfv\'{e}n times before the onset of the instability. This behaviour is observed for all the eruptive-runs, even if the Dispersion Central run shows a proportionally larger (smaller) decrease (increase) during the early (main) stage of the convergence phase. 

To summarize, the analysis shows that (1) apart from a single case ($\Hpj/\Phi^2$ for the Dispersion Central run) the time evolution of all the different helicity terms shows a difference between the eruptive and non eruptive runs, (2) this is not the case for the different magnetic energy terms,  where only $\Ej$ shows a different trend. (3) At the time of the onset of the eruption  a change in the trend is observed for all the simulation runs (apart from $\E$ for the Dispersion Central run). The fact that  this change in trend is also observed for the control, non-eruptive run suggests that the change in the coronal diffusion and in the boundary motions at the time of the eruption (see Section~\ref{Sec:MHDsim}) may play an important role.

\section{Thresholds in  magnetic energy and helicity}
\label{Sec:Thresholds}

In the previous Section we investigated the evolution of different magnetic energy and helicity related quantities around the moment of the onset of the torus instability, and we have discusses how a change in the trend of the different curves that occurs at $t\simeq t_I$, may be somehow related to these imposed boundary conditions. This is the reason why for any given quantity the existence of a threshold at the moment of the onset of the instability may be more important than a change in its trend. 

\citet{Zuccarello15} analyzed  these simulations  in the framework of the torus instability. In this framework, the instability occurs when the flux rope axis reaches a height where the decay index of the magnetic field has a critical value that depends on the particular magnetic field configuration. For these parametric simulations,  \citet{Zuccarello15} have shown that when an eruption occurs all the flux ropes have reached heights where the decay index $n$ has a critical threshold value of $n\simeq1.45\pm0.05$. 

The aim of this Section is to investigate if a critical threshold value in any of the different energy and helicity decompositions exists. Said differently, whether or not any of the  different energy and helicity decompositions have the same value (for all the simulations) when the instability sets in and the eruptions occur.  The values of the different quantities around the time of the onset of the  eruption are reported in Table~\ref{table:2}.

\subsection{Magnetic energy and helicity terms}
\label{Sec:ErupTerms}

Figure~\ref{Fig:E_shift} shows that no threshold in the total magnetic energy exists at the moment of the eruption. More specifically, Table~\ref{table:2} shows that $\E$ varies from about 163.7 for the Dispersion Peripheral run to about 206.5 for the Convergence run. The dispersion of the values, evaluated as $\left(\max \left[ \E \left(\bar{t}_I \right) \right]  - \min\left[ \E \left(\bar{t}_I \right) \right] \right)/ \max \left[ \E \left( \bar{t}_I \right) \right]$, is too large to correspond to a instability threshold solely based on that quantity. Indeed, if such threshold existed and corresponded to $ \max \left[\E \left(\bar{t}_I\right)\right]=206.5$ then no eruption should have been observed for the all the simulations but the Convergence run, since their $\E$ did not reach that threshold value. If the instability threshold was equal to $ \min \left[\E \left(\bar{t}_I\right)\right]=163.7$, then the eruption time, t$\simeq t_I$ should have been different, since all simulations but the Dispersion Central run would have reached that value of $\E$ earlier than their corresponding $t_I$.

For the type of numerical experiment presented here, in which the system is dynamically evolved from a stable to a unstable stage, the existence of a instability threshold uniquely based on a given quantity, $Q$, necessarily implies that the value of $Q(\bar{t}_I)$ should be the same for all the eruptive simulations. The measurement of the dispersion of $Q(\bar{t}_I)$ between the eruptive runs (as done above), is thus a way to state on the existence of a threshold for that quantity. The dispersion of about 20\% obtained for $\E$ disqualifies the existence of a threshold based on that quantity. 

A similar conclusion can be drawn also for $\Ej$ (Figure~\ref{Fig:Efree_shift}). At the moment of the onset of the eruption $\Ej$ has different values for the different runs with a range of dispersion of about 17\% of $\Ej$ of the Convergence run. Therefore, also $\Ej$ does not allow to distinguish the onset of the instability. The discrepancies are even larger when the potential magnetic energy, $\Ep$, is considered (Figure~\ref{Fig:Epot_shift}).

Figure~\ref{Fig:H_shift} (bottom panels) shows that no threshold exists also for the different decompositions of the normalized magnetic helicity. However, a closer inspection of the Figure and of Table~\ref{table:2} show that the dispersion of the different helicity and normalized-helicity terms is within 13\% (9\% for $\Hj$), i.e., the dispersion between the different helicity curves is about half the dispersion of the total energy curves.

\subsection{Current-carrying to total magnetic energy and helicity ratios}
\label{Sec:ErupRatios}

\cite{Pariat17} have shown that the ratio $\Ej/\E$  is a possible good eruption proxy, in the sense that it could discern between erupting and non-erupting runs. The same authors have shown that a significantly better proxy is the ratio of the helicity of the current-carrying part of the magnetic field to the total magnetic helicity, $\Hj/\Hv$, in the sense that this proxy has consistently larger values before the eruption for eruptive runs than for non-eruptive and, after the eruption, the proxies of eruptive and non-eruptive are indistinguishable.
 
\begin{figure}[!t]
\centering
\subfigure{	
\includegraphics[width=.4\textwidth,viewport= 7 18 552 558,clip] {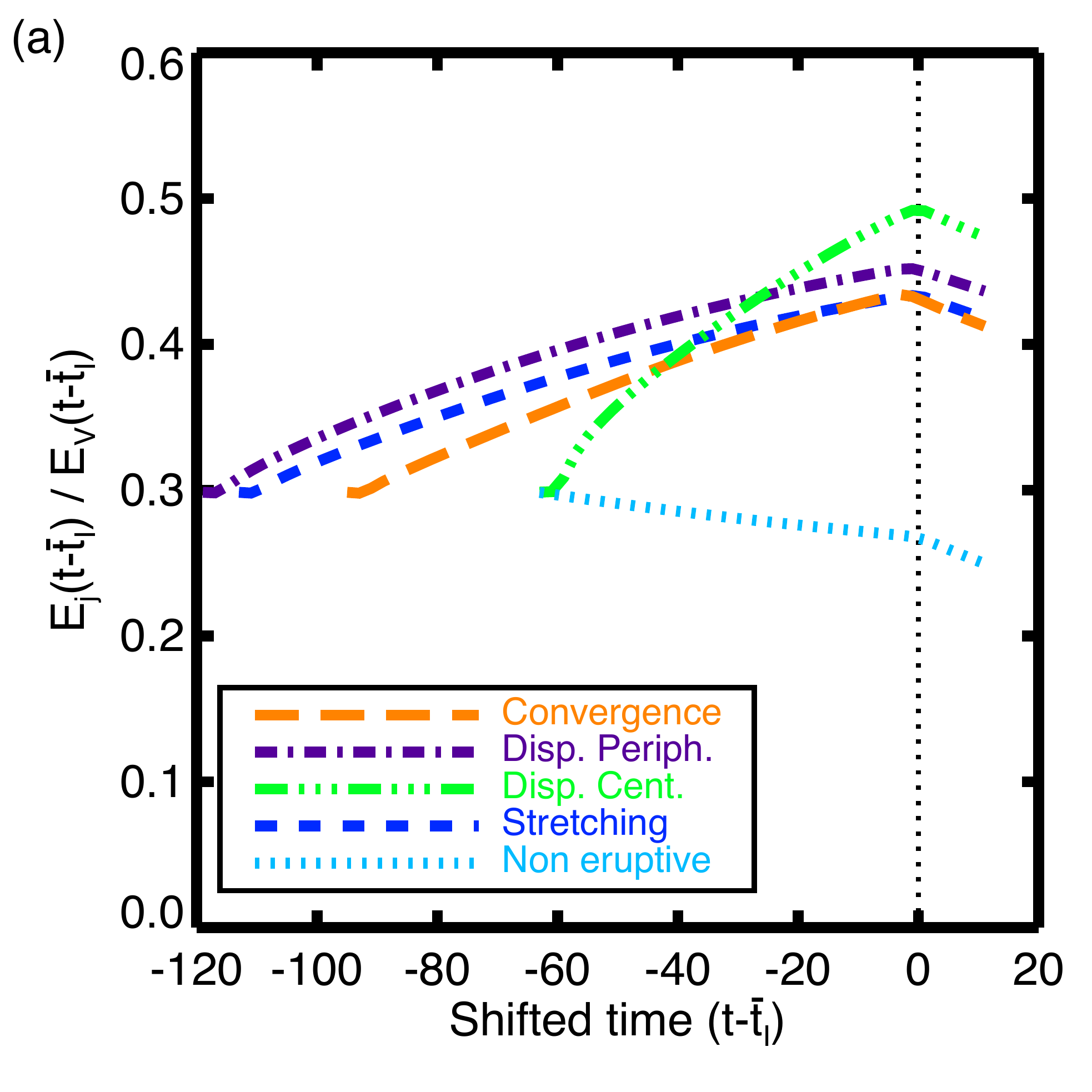}\label{Fig:Efree/Etot}
}
\subfigure{	
\includegraphics[width=.4\textwidth,viewport= 7 18 552 558,clip] {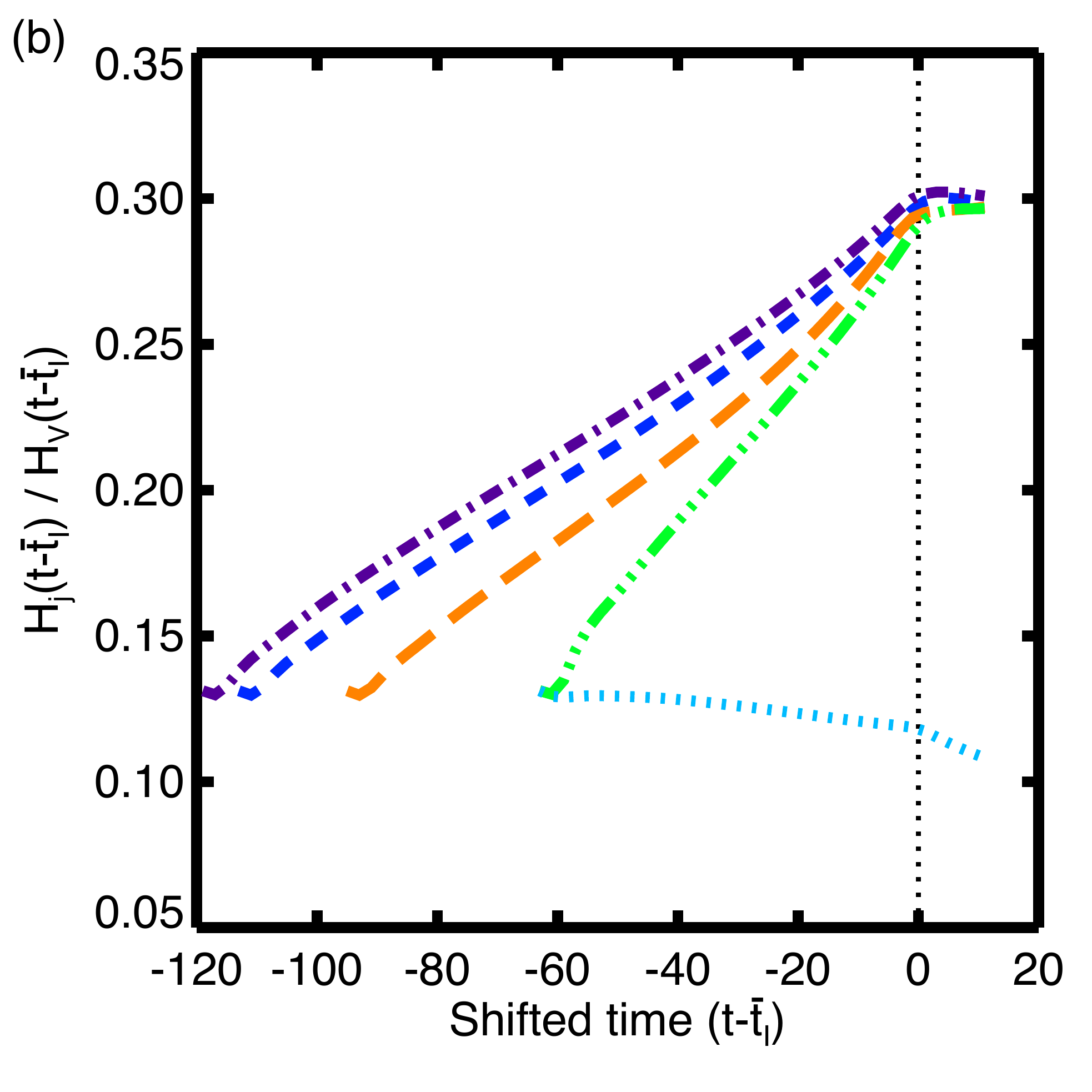}\label{Fig:Hj/Hv}
}
\caption{Time evolution of the $\Ej/ E$ and $\Hj/ \Hv$ ratios around the onset of the eruptions. \label{Fig:Ratios}}
\end{figure}

The time evolution of the  $\Ej/\E$  and $\Hj/\Hv$ for our simulations is shown Figure~\ref{Fig:Ratios}. At the moment of the eruption's onset  no threshold is observed in the  $\Ej / \E$  ratio (Figure~\ref{Fig:Efree/Etot}); the different values have a dispersion of about 10\% of the run with the highest value. 

The situation is significantly different when the $\Hj / \Hv$ ratio is considered. In fact, as shown in Figure~\ref{Fig:Hj/Hv} all the curves approach the same threshold value  within a dispersion of about 3\%. This dispersion range is between 8 and 3 times smaller than the equivalent ranges in the different energies and helicities decompositions discussed in the previous Section, and about 3 times smaller than $\Ej / \E$ ratio. We note that this dispersion is (1)  within the measurement precision of the helicity ratio, which is about 4\% (see Section~\ref{Sec:Helicity}), hence basically the same value, and (2) it is  about a factor 2 smaller than the dispersion of critical decay index values identified through the detailed analysis of the electric currents and magnetic field distribution in the different simulations \cite[see][]{Zuccarello15}.

%-----------------------------------------------------------------------------------
% Hj/Hv  Figure
%-----------------------------------------------------------------------------------

%-----------------------------------------------------------------------------------

\section{Discussion and Conclusion}
\label{Sec:Conclusion}

We have presented a series of eruptive and non-eruptive numerical MHD simulations of idealized solar active regions, which evolution is characterized by different boundary motions. With these series of simulations we aimed at addressing (1) which of the different boundary motions are the most efficient to inject different decompositions of magnetic energy and helicity, and (2) whether any of the different energy and helicity decompositions is able to identify the moment of the onset of the eruptions. 

The initial configuration consisted of an asymmetric, current-free, bipolar active region embedded in a constant Alfv\`{e}n speed atmosphere. 

During the first phase of the simulation runs, called shearing phase, shearing motions have been applied in the proximity of the active region's polarity inversion line (PIL). As a result, the coronal magnetic field evolves from a potential field into a current-carrying magnetic field characterized by a sheared arcade close to the PIL. 

Starting from this configuration four different classes of boundary motions, resembling motions often observed on the Sun, have been applied. This phase was called the flux rope formation phase. While the applied motions are relatively different among them, a characteristic that is common to all these four  motions is that they advect part of the photospheric magnetic flux towards the PIL. During this flux rope formation phase a change in the topology of the system is observed and a magnetic flux rope is formed, which eventually erupts  \citep{Zuccarello15}. 

By analysing the time evolution of the different magnetic energy and helicity decompositions during the sharing and flux rope formation phases we have shown that:

\begin{itemize}
\item Magnetic energy and total relative helicity are mostly injected during the shearing phase. The magnetic energy actually shows a significant decrease during the flux rope formation phase. This is due essentially to the decrease of the potential magnetic energy of the system, probably due the fact that magnetic flux is canceled at the PIL during the flux rope formation phase. 
 
\item Shearing motions are the most efficient to inject $\Ej$  into the system. The  injection rate of  $\Ej$ during this phase is at least four times larger than during the flux rope formation phases. 

\item The current-carrying component of the magnetic helicity, $\Hj$, increases with a similar rate between the shearing and the flux rope formation phases. 
\end{itemize}

In order to determine if any signature of the eruption's onset could be found in any of the magnetic energy or helicity decompositions, we analysed the evolution of these quantities around the moment of the onset of the eruption, to investigate if a threshold in any of these quantities exists.  Our analysis showed that: 

\begin{itemize}
\item No threshold is observed for any of the quantities entering in the decomposition of the magnetic energy (Eq.~\ref{eq:HDecomp}) and relative helicity (Eq.~\ref{eq:thomson}). In the different simulations the eruption occurs for various values of energies and helicities. The dispersion of these values are between 9\% and 25\% depending on the particular decomposition (with helicities decomposition in the lower part of this range). 
  
\item A threshold appears to exist in the ratio between the current-carrying component of the magnetic helicity and the total relative magnetic helicity. The onset of the eruptions indeed occurs when the different eruptive simulations reach the very same value of $\Hj/ \Hv$, within
measurement precision. This is not the case when a similar ratio in energies, i.e., $\Ej / E$, is considered.
\end{itemize}

\cite{Pariat17} have already discussed the promising properties of the ratio $\Hj/ \Hv$, as possible eruptivity proxy. The numerical experiments set-up was however limited in the sense that it could not conclude on the existence of a threshold since the magnetic system were not driven to instability in a controlled way from a stable configuration. This caveats is lifted for the numerical experiments analyzed in the present study.

For the same simulations discussed in this paper, \cite{Zuccarello15} performed a detailed analysis of the current distributions as well as several relaxation runs to determine the onset of the eruption. These authors have concluded that the driver of the eruption is indeed the torus instability, however, different simulations had a slightly different critical values of the decay index. They found that the critical value of the decay index at the onset of the eruptions is in the range $ n_{critical} \in [1.4,1.5]$. 

The torus instability occurs when the magnetic pressure of the current-currying flux rope is not balanced by the magnetic tension of the magnetic field `external' to it. The condition for the instability has been first derived analytically using infinitesimal current rings. It has been shown that the instability occurs when the apex of the current ring is in a location where the decay index is $n \gtrsim n_{critical} = 1.5$ \citep{Bateman78,Kliem06,Demoulin10}. There are several reasons why the critical decay index may differ among the different simulations: slightly different flux rope morphologies, limitations in determining the axis of non-analytical flux ropes and slightly different line-tying effects being the most relevant. Nevertheless, the clear result of \citet{Zuccarello15} was that when the flux rope's axis has reached an height where the decay index is $n \simeq 1.45 \pm 0.05$ a full eruption occurs. This value is remarkably close to the critical value for an idealized current ring. 

The helicity of the current-carrying component of the magnetic field, $\Hj$, is only related to the distribution of the electric currents. On the other hand, $\Hv$ also accounts for the contribution of the interaction  between this magnetic field and the potential field. The ratio $\Hj/ \Hv$ estimates the importance of $\Hj$ over $\Hv$, i.e., the importance of the field only associated with the currents over the total field. In our simulations, the torus instability occurs when the current-carrying flux rope has enough magnetic pressure that cannot be balanced by the tension of the potential field associated with the given boundary. For the present simulation set-up, when this occurs $\Hj$ is about one third of $\Hv$,  i.e., enough currents, associated with twisted, pressure-carrying magnetic fields, have been accumulated and they cannot be balanced any more by the tension of the potential field associated with the given boundary. 

In this paper we have shown that at the moment of the onset of the torus instability the ratio $H_j/H_V \simeq 0.29 \pm 0.01$ for four different simulations of torus unstable flux ropes. This suggests that the ratio $H_j/H_V$ is a good proxy for the onset of the torus instability, at least for this set of simulations. Some caution should however be taken in interpreting the particular value of $0.29$. Relative magnetic helicity, as defined in this study, is not a simply additive quantity. It implies that would the helicities have been computed in a different volume, using different boundary locations, a different value of the helicity threshold may have been obtained. In the present study the values obtained between the simulations are consistent with each other, because they are computed on the very same numerical domain,  which robustly validate the core results of the existence of a threshold on $H_j/H_V$ for these simulations. The specific value obtained is however likely not universal. This value should not be taken straightforwardly as an eruption trigger criteria in, for example, observational studies, before further studies have been carried out. Relative magnetic helicity remains a poorly understood physical quantity which may need to be theoretically partly redefined and which properties need to be further understood \citep[e.g. as in][]{Demoulin06,Yeates13,Dalmasse14,Dalmasse18,Russell15,Aly18,Oberti18}.

If the ratio $H_j/H_V$ turns out to be either physically related to the torus instability or just a good proxy of it, this would constitute a significant step forward in forecasting solar eruptions. In fact, the determination of the eruptivity potential of an active region based on the evaluation of the decay index can be achieved through observations \citep{Kliem13,Zuccarello14,Zuccarello16, James18}. However, its routine application might not be straightforward: it requires to address non trivial problems such as defining and identifying the axis of non-analytical flux ropes in strongly asymmetric configurations and inferring the three-dimensional nature of solar filaments from stereoscopic observations. On the other hand, evaluating the ratio $H_j/H_V$ would ‘only’ require the construction of a three-dimensional magnetic field model of the active region, which automation could be achieved more easily than the other approach. However, before considering all of the above, the robustness of this criterion, whether or not the threshold effectively exists, and if its value is magnetic system independent, needs to be extensively tested, first using as many numerical experiments as possible, and then against observed data.

\begin{acknowledgements}
F.~P.~Z. acknowledges the support of the Research Foundation - Flanders, FWO grant no. 1272718N. 
E.~P. acknowledges the support of the FLARECAST project, funded by the European Union’s Horizon2020 research and innovation program under grant agreement n$^\circ$ 640216.
E.~P. \& L.~L. acknowledge the support of the French Agence Nationale pour la Recherche  through the HELISOL project, contract n$^\circ$ ANR-15-CE31-0001.
G.~V. acknowledges the support of the Leverhulme Trust Research Project Grant 2014-051.
This work was granted access to the HPC resources of MesoPSL financed by the R\'{e}gion Ile de France and the project Equip@Meso (reference ANR-10-EQPX-29-01) of the programme Investissements d' Avenir supervised by the Agence Nationale pour la Recherche. 
  
\end{acknowledgements}

\end{document}